\RequirePackage{fix-cm}
\documentclass[smallextended]{svjour3}       
\smartqed  
\usepackage{graphicx}
\journalname{Journal of Elasticity}
\begin{document}

\title{Large-Amplitude Elastic Free-Surface Waves: \\
Geometric Nonlinearity and Peakons
}

\titlerunning{Nonlinear Surface Waves}        

\author{Lawrence K. Forbes \and Stephen J. Walters
        \and Anya M. Reading
}

\authorrunning{L.K. Forbes, S.J. Walters, A.M. Reading} 

\institute{Lawrence K. Forbes\at
              School of Mathematics and Physics \\
							University of Tasmania, Hobart Tasmania \\
              Tel.: +61-03-62262720\\
              Fax: +61-03-62262410\\
              \email{larry.forbes@utas.edu.au}        \\
              \emph{ORCID number:} 0000-0002-9135-3594  
           \and
           Stephen J. Walters \at
              School of Mathematics and Physics \\
							University of Tasmania, Hobart Tasmania \\
						\and
						Anya M. Reading \at
							School of Mathematics and Physics \\
							University of Tasmania, Hobart Tasmania
}


\maketitle

\begin{abstract}
An instantaneous sub-surface disturbance in a two-dimensional elastic half-space 
is considered.  The disturbance propagates through the elastic material until it reaches 
the free surface, after which it propagates out along the surface.  In conventional theory,
the free-surface conditions on the unknown surface are projected onto the flat plane
$y = 0$, so that a linear model may be used.  Here, however, we present a formulation that
takes explicit account of the fact that the location of the free surface is unknown
{\it a priori}, and we show how to solve this more difficult problem numerically.  This 
reveals that, while conventional linearized theory gives an accurate account of the 
decaying waves that travel outwards along the surface, it can under-estimate the 
strength of the elastic rebound above the location of the disturbance.  In some 
circumstances, the non-linear solution fails in finite time, due to the formation of 
a ``peakon'' at the free surface.  We suggest that brittle failure of the elastic 
material might in practice be initiated at those times and locations.
\end{abstract}

\section{Introduction}
\label{intro}

This paper is concerned with the propagation of large-amplitude waves along the
free surface of a semi-infinite elastic medium.  Novel physical phenomena, and
particularly those associated with nonlinear effects at the free surface, are of
interest here.  Accordingly, in this work we study a finite-amplitude version of 
the classical linearized problem introduced by Lamb.  We consider 
an idealized situation in which an impulsive disturbance to the
velocity at some submergence depth $H$ occurs at the initial time $t = 0$, 
after which waves move through the medium and generate finite-amplitude 
free-surface waves that move outwards from the disturbance.

The propagation of small-amplitude waves at a free surface on a semi-infinite 
medium is a classical problem in elasticity, resulting in Rayleigh, Love 
and Lamb waves.  Overviews are given in the solid Earth context by 
Aki and Richards \cite{AkiRichards}, and for engineering applications 
by Worden \cite{Worden}.  The problem considered here, in which 
waves are excited at the free surface by a submerged disturbance, was
first solved by Lamb for weak disturbances of Dirac delta-function type; this
is now known as ``Lamb's Problem'' and is analyzed in detail in the text 
\cite[Chapter 6]{AkiRichards}.  A slight simplification of this difficult,
but linear, problem is solved by Kausel \cite{Kausel}.

In these classical analyses, the governing equations are \textit{linearized},
which is appropriate for small-amplitude disturbances.  This requires two
different mathematical approximations, and corresponds to two separate
physical assumptions about the material.  The first, and most obvious, is to
overlook any effects of \textit{material nonlinearity} and to suppose, therefore,
that the stress and strain tensors in the elastic solid are linearly related.
This essentially implies that Hooke's Law holds throughout the material
\cite[Section 2.2]{AkiRichards}.  

The second step in linearization is to ignore the effects of 
\textit{geometric nonlinearity}, and this is a more subtle and nuanced form of 
approximation.  The initial location of the free surface may originally have been
simply the plane $y = 0$, but this ceases to be true once the material starts to
deform.  Some time later, the free surface will have some unknown location 
$y = S \left( x,t \right)$.  A consequence is that physical conditions
to be applied at the free surface can no longer be supposed to hold simply on
the plane $y = 0$, but rather must now be applied on the actual unknown location
$y = S$ itself.  

In most studies of elastic free-surface waves, it is simply assumed that any
conditions to be applied at the actual free surface can instead be imposed just 
at the undisturbed level $y = 0$.  In addition, any nonlinear terms in the
surface function $S$ or its derivatives are ignored.  It is easy enough, on the
basis of straightforward perturbation theory, to argue that making these 
approximations is consistent with assuming that the surface waves are of small
amplitude and contain no steep sections in their profiles.  For elastic free-surface
waves, the boundary condition is that the normal and the tangential components 
of the traction vector are both zero at the free surface, and when geometric
nonlinearity is ignored, a linearized approximation to these conditions is
assumed to hold simply on the undisturbed location $y = 0$ of the surface 
(see \textit{e.g.} Lan and Zhang \cite{LanZhang}, their equations 3 and 4).

Nevertheless, once the finite deformation of the elastic medium is taken into
account, the displacement of the free surface from its undisturbed level $y = 0$
can no longer be overlooked.  To account for this effect, several different approaches
have been considered in the literature.  In the first of these, the (possibly linear)
equations of elasticity in the medium are combined with the nonlinear kinematic
conditions at the free surface, and then averaged over the material to some order
of approximation.  This yields a weakly nonlinear partial differential equation (PDE)
for the vertical displacement, and such an equation has been considered by
Lenells \cite{Lenells}.  His equation was modified slightly by Xie \textit{et al.}
\cite{XieEtAl} to include a type of damping term.  Those authors focussed on the
presence of ``peakons'' in their solutions, which are periodic travelling waves with 
sharp peaks either at their crests or their troughs, depending on the values of certain
parameters.  They used bifurcation analysis in a phase plane to show how these
peakon solutions arise from the nonlinearity of their (integrable) equation, through
heteroclinic orbits with three-fold symmetry.  Discontinuous and peaked waveforms
had also been obtained earlier, using numerical integration, by Lardner \cite{Lardner}.

A second approach to retaining the effects of geometric nonlinearity at the 
free surface has been suggested more recently by Pal \textit{et al.} \cite{PalEtAl}.  
They simulated a continuum using a hexagonal lattice of Hookean springs.  This is an
illustrative example of pure geometric nonlinearity since, while the springs have a linear
stress-strain behaviour when extended along their length, their motion also includes
displacements at varying lateral angles, which are nonlinear functions of their 
extensions.  (In fact, even a simple system consisting of only two Hookean springs
can give rise to highly nonlinear behaviour when the motion occurs transversely
\cite{Forbes1991}; see also \cite{Amabili}).

Perturbation methods have also been used to explore the effects of geometric nonlinearity,
and some of these are reviewed by Maugin \cite{Maugin2007}, who additionally illustrates a
calculation of an elastic free-surface wave to second order in the wave amplitude by means
of a method of strained coordinates.  Recently, Wang and Fu \cite{WangAndFu} have used 
a perturbation method to third order in amplitude to impose the traction-free boundary
condition on the exact wrinkled free surface of a half-space elastic body subject to
lateral compression.  Their analysis is lengthy and involved, but does allow them to
determine conditions under which the compressed surface will wrinkle.  Finally,
the effects of geometric nonlinearity at the free surface have been studied directly,
using numerical methods.  Clayton and Knap \cite{ClaytonKnap} formulate their
problem using phase type methods to cope with the location of the unknown surface,
and this enables them to follow the progress of a crack developing in the material.
Nonlinear solutions of soliton type have been obtained by Pouget and Maugin 
\cite{PougetMaugin1}, \cite{PougetMaugin2}.  Their model derives from an assumption 
that the elastic behaviour itself remains linear but that there are large-amplitude 
distortions present in the material; this therefore corresponds to geometric nonlinearity
rather than material nonlinearity.

When nonlinearities of both forms are ignored, small-amplitude elastic free-surface
waves are obtained as the solution of the linear partial differential equations
of momentum conservation in the half-space $y < 0$, with linear boundary conditions
imposed on the surface $y = 0$, as discussed above.  In principle, since these are
linear equations with constant coefficients, their solution can be done in a
classical manner, using a combination of Fourier and Laplace Transforms.  However, 
inverting those Transforms to obtain the final free-surface shapes is not straightforward,
and in practice the result is obtained only as a difficult integral that must be
evaluated numerically.  From a physical point of view, this well-known difficulty
with the linear two-dimensional elasticity equations is due in part to the fact that 
they contain two different wave speeds, corresponding to S and P waves, as discussed by 
Ockendon \& Ockendon \cite{OckOck}, and that energy can transfer back and forth
between these shear and compressive transmission modes.  When no free surface is
present and the waves can simply move through an unbounded medium, closed-form
mathematical solutions are occasionally possible, and some of these have been presented
by Walters \textit{et al.} \cite{Walters2020}.  The analysis is very greatly complicated
by the presence of a free surface, however, and can involve sophisticated manipulation
of integrals in the complex Laplace-Transform space, containing awkward branch lines.
Further discussion of these techniques is presented by Aki and Richards 
\cite[Sections 6.4, 6.5]{AkiRichards} as well as the two papers of Diaz and Ezziani
\cite{Diaz2D}, \cite{Diaz3D}.

On the other hand, these linearized elastic free-surface waves can be computed
accurately and rapidly using numerical finite-difference techniques.  This is due
to the fact that these are wave equations (hyperbolic PDEs).  Lan and Zhang 
\cite{LanZhang} discuss various finite-difference schemes when the linearized
free-surface conditions are imposed on $y = 0$, and an alternative discussion of
finite-difference methods is given by Min {\it et al}. \cite{MinEtAl}.  Here, we 
present a straightforward explicit finite-difference method for computing linearized
free-surface waves in Section \ref{sec:explicit}, and then develop a more robust
implicit ADI scheme for these waves in Section \ref{sec:implicit}.  

This new implicit scheme then forms the backbone of a numerical method that allows us
to explore in detail the effects of geometric nonlinearity on an idealized problem, in
which a buried disturbance in an elastic half space generates waves that first move 
outward and upward from the disturbance until they interact with a free surface, 
where they trigger surface waves that move laterally from the original site 
of the disruption.  This is related to the Lamb Problem \cite{Kausel}, but here, 
we take into account the finite movement of the surface itself.  This results in 
a highly non-linear mathematical problem, and we find that these finite-amplitude 
nonlinear elastic free-surface waves introduce two new phenomena into the physics, 
for which there is no linearized counterpart.  Firstly, there can be a strong 
rebound near the initial disturbance.  Secondly, a surface-wave singularity 
in the curvature may be produced within finite time, for sufficiently intense 
initial disturbances.  This manifests as a discontinuity in the derivative of
the surface profile at a wave peak, and so has been labelled a ``peakon'' by
some authors using weakly nonlinear asymptotic theories to study such phenomena
\cite{XieEtAl}, \cite{Salupere}.  We discuss a possible physical meaning 
for these peakon curvature singularities in Section \ref{sec:conclude}.

\section{Governing Equations}
\label{sec:governing}

We consider an elastic solid in two-dimensional geometry, occupying the lower
half-plane $y < 0$.  At the initial time $t = 0$, the free surface is simply the
horizontal plane $y = 0$.  Body forces on the material are ignored, and the solid
has density $\rho$ and Lam{\' e} parameters $\lambda$ and $\mu$, which are taken
in this study simply to be constants throughout the material.  A disturbance
centred at $\left( x,y \right) = \left( 0 , - H \right)$ below the surface
is activated at $t = 0$, and causes waves to propagate through the medium and
along the free surface. 

The displacement vector ${\bf u}$ for particles of the solid is represented 
in cartesian coordinates as ${\bf u} = u^X {\bf i} + u^Y {\bf j}$ for 
two-dimensional geometry, in which the symbols ${\bf i}$ and ${\bf j}$ denote 
unit vectors in the positive $x$- and $y$-directions, respectively.  Conservation 
of linear momentum is expressed through the Cauchy momentum equation which, 
when combined with generalized Hooke's Law, gives rise to the elastodynamic 
momentum equations
\begin{eqnarray}
\frac {\partial^2 {\bf u}}{\partial t^2} 
& = & \frac {\lambda + \mu}{\rho} \nabla \left( \nabla\cdot {\bf u} \right)
+ \frac {\mu}{\rho} \nabla^2 {\bf u}   .
\label{eq:Hooke}
\end{eqnarray}
This system of partial differential equations (\ref{eq:Hooke}) is famously difficult to
solve in closed form (see Ockendon \& Ockendon \cite[section 3.8.3]{OckOck}), despite 
being linear and having constant coefficients.  One reason for this is that its solution 
vector ${\bf u}$ can be decomposed into the sum of the gradient of a scalar potential and
the curl of a vector potential, and both potentials satisfy separate wave equations,
one with speed $c_P = \sqrt{ \left( \lambda + 2\mu \right) / \rho}$ corresponding
to a compression wave, and the other with the slower speed $c_S = \sqrt{ \mu / \rho}$
of a shear wave.  Since the system is elastic, energy is able to be transferred
between the compression and shear modes of propagation (\cite[page 76]{OckOck}).

It is convenient now to non-dimensionalize the problem, using the submergence depth
$H$ of the disturbance as the length scale, and the shear-wave speed 
$c_S = \sqrt{ \mu / \rho}$ as the unit of speed.  Accordingly, time $t$ is made
dimensionless with respect to the reference time $H \sqrt{ \rho / \mu}$.  
Non-dimensional variables will be used henceforth.

The elastodynamic equation (\ref{eq:Hooke}) now becomes
\begin{eqnarray}
\frac {\partial^2 u^X}{\partial t^2} & = & \alpha^2 \frac {\partial^2 u^X}{\partial x^2}
+ \left( \alpha^2 - 1 \right) \frac {\partial^2 u^Y}{\partial x \partial y}
+ \frac {\partial^2 u^X}{\partial y^2} ,   
\nonumber   \\
\frac {\partial^2 u^Y}{\partial t^2} & = & \alpha^2 \frac {\partial^2 u^Y}{\partial y^2}
+ \left( \alpha^2 - 1 \right) \frac {\partial^2 u^X}{\partial x \partial y}
+ \frac {\partial^2 u^Y}{\partial x^2} .
\label{eq:HookeComponent}
\end{eqnarray}
This dimensionless system (\ref{eq:HookeComponent}) involves only the single non-dimensional 
parameter
\begin{equation}
\alpha^2 = \left( \lambda + 2 \mu \right) / \mu   ,
\label{eq:ParamA}
\end{equation}
which may be regarded as $c_P^2 / c_S^2$ , the ratio of the squares of the speeds of
the compressive and shear waves.  The two Lam{\' e} parameters $\lambda$ and
$\mu$ are assumed here to be positive, so then  $\alpha^2 > 2$.

We regard the free surface of the elastic material as some curve 
$y = S \left( x,t \right)$ of unknown location.  Initially, 
$S \left( x,0 \right) = 0$, and the surface then evolves with time.  However, the
free surface is a material boundary, and so in this Lagrangian description,
in which $\left( x(t) , y(t) \right)$ represents the position vector of a
particle on the surface, it then follows that  $y(t) = S \left( x(t) , t \right)$.
It is convenient in this analysis to differentiate with respect to time following
the material surface-particle, and so to obtain the kinematic boundary condition
in the form
\begin{equation}
\frac {\partial S}{\partial t} = \frac {\partial u^Y}{\partial t}
- \frac {\partial S}{\partial x} \frac {\partial u^X}{\partial t}
\quad \textrm{on } y = S \left( x,t \right)  .
\label{eq:Kinem}
\end{equation}
The dynamic condition at the free surface is that both the normal and the
tangential stress components must be zero there.  If the stress tensor in 
the solid is written ${\bf T}$, then the dynamic conditions are
\begin{equation}
{\bf \hat{n} \cdot T \cdot \hat{n}} = 0 \quad \textrm{;} \quad
{\bf \hat{t} \cdot T \cdot \hat{n}} = 0
\quad \textrm{on } y = S \left( x,t \right)  ,
\label{eq:Dynamic}
\end{equation}
in which the two unit vectors normal and tangential to the free surface are
\begin{equation}
{\bf \hat{n}} = \left( - S_x {\bf i} + {\bf j} \right) / \sqrt{ 1 + S_x^2} 
\quad \textrm{;} \quad
{\bf \hat{t}} = \left( {\bf i} + S_x {\bf j} \right) / \sqrt{ 1 + S_x^2}
\label{eq:SurfaceVectors}
\end{equation}
and the subscript denotes partial differentiation with respect to the indicated
variable $x$.  After some algebra, the first of the equations in (\ref{eq:Dynamic}) 
gives the normal dynamic free-surface condition
\begin{eqnarray}
& & S_x^2 \left[ \alpha^2 \frac {\partial u^X}{\partial x}
+ \left( \alpha^2 - 2 \right) \frac {\partial u^Y}{\partial y} \right]
- 2 S_x \left[ \frac {\partial u^X}{\partial y} 
+ \frac {\partial u^Y}{\partial x} \right] 
\nonumber   \\
& & + \left[ \left( \alpha^2 - 2 \right) \frac {\partial u^X}{\partial x}
+ \alpha^2 \frac {\partial u^Y}{\partial y} \right] = 0
\nonumber   \\
& &  \quad \textrm{on } y = S \left( x,t \right)  .
\label{eq:DynamicNormal}
\end{eqnarray}
Similarly, the second equation in the system (\ref{eq:Dynamic}), with the 
results (\ref{eq:SurfaceVectors}), gives the tangential dynamic condition
\begin{eqnarray}
& &  2 S_x \left[ \frac {\partial u^Y}{\partial y} 
- \frac {\partial u^X}{\partial x} \right]
+ \left( 1 - S_x^2 \right) \left[ \frac {\partial u^X}{\partial y}
+ \frac {\partial u^Y}{\partial x} \right] = 0
\nonumber   \\
& &  \quad \textrm{on } y = S \left( x,t \right)  .
\label{eq:DynamicTangent}
\end{eqnarray}

Finally, initial conditions are needed, to close the problem.  For definiteness
we assume that the initial displacements are all zero, but that an impulsive 
initial velocity disturbance, centred at the point 
$\left( x,y \right) = \left( 0, -1 \right)$, is made to the system. Here we
suppose that
\begin{equation}
 u^X \left( x,y,0 \right) = 0 \quad \textrm{;} \quad u^Y \left( x,y,0 \right) = 0 
\quad \textrm{;} \quad S \left( x,0 \right) = 0 ,
\label{eq:InitialDisp}
\end{equation}
but that
\begin{eqnarray}
\frac {\partial u^X}{\partial t} \left( x,y,0 \right) & = & 
K_A \frac{2x}{a^2} \exp \left[ - \left( \frac {x}{a} \right)^2
- \left( \frac {y + 1}{b} \right)^2 \right] ,   
\nonumber   \\
\frac {\partial u^Y}{\partial t} \left( x,y,0 \right) & = &
K_A \frac{2 \left( y + 1 \right)}{b^2} \exp \left[ - \left( \frac {x}{a} \right)^2
- \left( \frac {y + 1}{b} \right)^2 \right] .
\nonumber   \\
& &   \label{eq:InitialVeloc}
\end{eqnarray}
The constants $a$ and $b$ determine the effective physical size of the initial
disturbance, and $K_A$ is its amplitude.

This system (\ref{eq:HookeComponent}), (\ref{eq:Kinem}), 
(\ref{eq:DynamicNormal})--(\ref{eq:InitialVeloc}) is highly nonlinear since, 
although the governing partial differential equations (\ref{eq:HookeComponent}) 
are linear, they are nevertheless to be solved in a region $y < S \left( x,t \right)$ 
of unknown shape determined by the nonlinear conditions (\ref{eq:Kinem}), 
(\ref{eq:DynamicNormal}) and (\ref{eq:DynamicTangent}).

\section{Linearized System}
\label{sec:linear}

If the amplitude $K_A$ of the initial disturbance is small, then it is to be
expected that the amplitude of the waves so produced should likewise be small.
In that case, the two displacements  $u^X$ and $u^Y$  and the free-surface
deformation $S$ can be expanded as power series in $K_A$, of the form
\begin{eqnarray}
u^X \left( x,y,t \right) 
& = & 0 + K_A U^{1X} \left( x,y,t \right) + {\cal O} \left( K_A^2 \right)   
\nonumber   \\
u^Y \left( x,y,t \right) 
& = & 0 + K_A U^{1Y} \left( x,y,t \right) + {\cal O} \left( K_A^2 \right)   
\nonumber   \\
S \left( x,t \right) 
& = & 0 + K_A S^{1} \left( x,t \right) + {\cal O} \left( K_A^2 \right) .
\label{eq:LinExpand}
\end{eqnarray}
These expansions (\ref{eq:LinExpand}) are substituted into the governing equations in
Section \ref{sec:governing} and terms are retained only to order $K_A$.

Since the elastodynamic equations (\ref{eq:HookeComponent}) are already linear, then the
first-order perturbation functions $U^{1X}$ and $U^{1Y}$ satisfy the same
system of partial differential equations (\ref{eq:HookeComponent}) as $u^X$ and $u^Y$.
However, the kinematic condition (\ref{eq:Kinem}) linearizes simply to 
\begin{equation}
S^{1} \left( x,t \right) = U^{1Y} \left( x, 0, t \right)
\label{eq:LinKinem}
\end{equation}
projected from the exact interface $y = S \left( x,t \right)$ to the
approximating plane $y = 0$.  The normal dynamic free-surface condition 
(\ref{eq:DynamicNormal}) now becomes approximately
\begin{equation}
\left( \alpha^2 - 2 \right) \frac {\partial U^{1X}}{\partial x}
+ \alpha^2 \frac {\partial U^{1Y}}{\partial y} = 0
\quad \textrm{on } y = 0  ,
\label{eq:LinDynNormal}
\end{equation}
and the tangential dynamic surface condition (\ref{eq:DynamicTangent}) linearizes to 
\begin{equation}
\frac {\partial U^{1X}}{\partial y}
+ \frac {\partial U^{1Y}}{\partial x} = 0
\quad \textrm{on } y = 0  .
\label{eq:LinDynTangent}
\end{equation}
These two linearized forms (\ref{eq:LinDynNormal}), (\ref{eq:LinDynTangent}) of 
the free-surface conditions are the ones usually found in investigations of propagating 
surface waves, such as those by Lan and Zhang \cite{LanZhang} and 
Min \textit{et al.} \cite{MinEtAl}.

In principle, the process for obtaining an exact solution to this linearized 
problem (\ref{eq:HookeComponent}), with boundary conditions 
(\ref{eq:LinKinem})--(\ref{eq:LinDynTangent}) and subject to initial conditions 
of the sort described by (\ref{eq:InitialDisp}), (\ref{eq:InitialVeloc}), should be 
reasonably straightforward.  Similar to the method detailed in Walters \textit{et al.}
\cite{Walters2020}, a Fourier Transform is taken in the spatial variable $x$, followed by 
a Laplace Transform in the time variable $t$.  A fourth-order differential equation 
in the variable $y$ is then solved, to give the complete solution in the Fourier-Laplace 
space.  Such a process is used by Diaz \& Ezziani \cite{Diaz2D}, \cite{Diaz3D} in 
their investigations of a related elastic-acoustic problem, for example.  The difficulty 
comes here, however, when attempting to invert the Laplace Transform, since the two 
different wave-speeds, corresponding to the compression and shear waves, give rise 
in the Fourier-Laplace space to difficult overlapping branch singularities in the 
complex domain of the transform variables.  Consequently, the transforms are extremely 
difficult to invert, and even when successful, the final answer is often only available 
in the form of a complicated integral that must be evaluated numerically.  Closed-form 
expressions for the solution variables are thus not easily accessible.

\subsection{Explicit Method}
\label{sec:explicit}

By contrast with analytical approaches, the numerical solution of this linearized 
system by finite differences is not difficult, runs quickly and gives accurate 
results.  The $x-y$ lower-half plane is approximated by the finite computational 
domain $-L < x < L$, $-H < y < 0$ and a uniform numerical grid
$x_1 , x_2 , \dots , x_M$ , $y_1 , y_2 , \dots , y_N$ is placed over the
region, such that $x_1 = -L$ and $x_M = L$, deep within the elastic 
material $y_1 = -H$ and on the linearized free surface $y_N = 0$.
We use the notation
\begin{equation}
X_{i,j}^k \equiv U^{1X} \left( x_i , y_j , t_k \right)
\quad \textrm{;} \quad
Y_{i,j}^k \equiv U^{1Y} \left( x_i , y_j , t_k \right)
\label{eq:LinDiscreteDisp}
\end{equation}
to represent discrete values of the two components of the linearized
displacement vector, and define the four difference operators
\begin{eqnarray}
\delta_x^2 X_{i,j}^k & = & \left( X_{i+1,j}^k - 2 X_{i,j}^k
+ X_{i-1,j}^k \right) / \Delta x^2   
\nonumber   \\
\delta_y^2 X_{i,j}^k & = & \left( X_{i,j+1}^k - 2 X_{i,j}^k
+ X_{i,j-1}^k \right) / \Delta y^2   
\nonumber   \\
\delta_t^2 X_{i,j}^k & = & \left( X_{i,j}^{k+1} - 2 X_{i,j}^k
+ X_{i,j}^{k-1} \right) / \Delta t^2   
\nonumber   \\
\delta_{xy}^2 X_{i,j}^k & = & \bigl( X_{i+1,j+1}^k - X_{i-1,j+1}^k
\nonumber   \\
& & - X_{i+1,j-1}^k + X_{i-1,j-1}^k \bigr) / \left( 4 \Delta x \Delta y \right)  .
\label{eq:LinDiffOperators}
\end{eqnarray}
In these expressions, the quantities $\Delta x$, $\Delta y$ and $\Delta t$
are the (uniform) spacings of the mesh points in the $x$, $y$ and $t$
coordinates, respectively.  

It is now relatively straightforward to create a simple explicit 
finite-difference scheme of second-order accuracy, by approximating the 
elastodynamic equations (\ref{eq:HookeComponent}) as
\begin{eqnarray}
\delta_t^2 X_{i,j}^k & = & \alpha^2 \delta_x^2 X_{i,j}^k
+ \left( \alpha^2 - 1 \right) \delta_{xy}^2 Y_{i,j}^k + \delta_y^2 X_{i,j}^k ,   
\nonumber   \\
\delta_t^2 Y_{i,j}^k & = & \alpha^2 \delta_y^2 Y_{i,j}^k
+ \left( \alpha^2 - 1 \right) \delta_{xy}^2 X_{i,j}^k + \delta_x^2 Y_{i,j}^k ,   
\nonumber   \\
& & i = 2, \dots , M-1  \quad \textrm{,} \quad j = 2, \dots , N  .
\label{eq:LinHookeExplicit}
\end{eqnarray}
On the computational boundaries $x_1$ , $x_N$ and $y_1$ the displacements
are simply set to zero.  

To cope with the dynamic boundary conditions, a ``false boundary'' 
$y \equiv y_{N+1} = \Delta y$ is introduced above the surface $y_N = 0$, 
and in fact, since the equations (\ref{eq:LinHookeExplicit}) are applied at the surface, 
where $j = N$, they already involve values on the false boundary $j = N+1$.  
These are eliminated using the tangential dynamic condition (\ref{eq:LinDynTangent})
\begin{eqnarray}
X_{i,N+1}^k & = & X_{i,N-1}^k - \frac {\Delta y}{\Delta x} \left( 
Y_{i+1,N}^k - Y_{i-1,N}^k \right) 
\nonumber   \\
& & \quad i = 2, \dots , M-1  ,
\label{eq:LinTangentExplicit}
\end{eqnarray}
and the linearized normal condition (\ref{eq:LinDynNormal})
\begin{eqnarray}
Y_{i,N+1}^k & = & Y_{i,N-1}^k - \left( \frac {\alpha^2 - 2}{\alpha^2} \right)
\frac {\Delta y}{\Delta x} \left( X_{i+1,N}^k - X_{i-1,N}^k \right) 
\nonumber   \\
& & \quad i = 2, \dots , M-1  .
\label{eq:LinNormalExplicit}
\end{eqnarray}

This system of difference equations (\ref{eq:LinHookeExplicit})--(\ref{eq:LinNormalExplicit}) is 
now able to be iterated forward in the index $k$ that corresponds to increasing time.  The 
initial conditions (\ref{eq:InitialDisp}) give
\begin{equation}
X_{i,j}^1 = 0 \quad \textrm{;} \quad Y_{i,j}^1 = 0 ,
\label{eq:LinInitExplicitDisp}
\end{equation}
and the derivative initial conditions (\ref{eq:InitialVeloc}) yield
\begin{eqnarray}
X_{i,j}^2 & = & X_{i,j}^1 
+ \Delta t K_A \frac{2x_i}{a^2} \exp \left[ - \left( \frac {x_i}{a} \right)^2
- \left( \frac {y_j + 1}{b} \right)^2 \right] ,   
\nonumber    \\
Y_{i,j}^2 & = & Y_{i,j}^1 
+ \Delta t K_A \frac{2 \left( y_j + 1 \right)}{b^2} 
\exp \left[ - \left( \frac {x_i}{a} \right)^2
- \left( \frac {y_j + 1}{b} \right)^2 \right] .
\nonumber   \\
& &   \label{eq:LinInitExplicitVeloc}
\end{eqnarray}
The difference scheme (\ref{eq:LinHookeExplicit}) is re-arranged to give explicit formulae 
for $X_{i,j}^{k+1}$ and $Y_{i,j}^{k+1}$ and iterated forward from the two
starting levels $k = 1$ and $k = 2$ provided by the initial conditions
(\ref{eq:LinInitExplicitDisp}) and (\ref{eq:LinInitExplicitVeloc}).  This works well, runs 
very quickly and gives results of good accuracy for small values of the speed-ratio 
parameter $\alpha^2 > 2$.  It is the method of choice for the results obtained 
with $\alpha^2 = 3$ to be presented in Section \ref{sec:results}.  However, as $\alpha^2$ 
increases, the numerical stability of this simple explicit scheme requires prohibitively
small values of the time step $\Delta t$, and an implicit scheme is needed
instead.  We have found that the implicit scheme is mandatory for the
non-linear solution in Section \ref{sec:nonlinear}, and so it is instructive
to demonstrate it now, for the simpler linearized problem.

\subsection{Implicit Method}
\label{sec:implicit}

We consider the same finite-difference grid described in Section \ref{sec:explicit}, 
and to derive an implicit scheme, we adapt the method developed by Qin \cite{Qin}.  
The elastodynamic equations (\ref{eq:HookeComponent}) (which are identical for 
the linearized displacements $U^{1X}$ and $U^{1Y}$) are now written as
\begin{eqnarray}
\frac {\partial v^X}{\partial t} & = & \alpha^2 \frac {\partial^2 u^X}{\partial x^2}
+ \left( \alpha^2 - 1 \right) \frac {\partial^2 u^Y}{\partial x \partial y}
+ \frac {\partial^2 u^X}{\partial y^2} ,   
\nonumber   \\
\frac {\partial u^X}{\partial t} & = & v^X ,   
\nonumber   \\ 
\frac {\partial v^Y}{\partial t} & = & \alpha^2 \frac {\partial^2 u^Y}{\partial y^2}
+ \left( \alpha^2 - 1 \right) \frac {\partial^2 u^X}{\partial x \partial y}
+ \frac {\partial^2 u^Y}{\partial x^2} ,   
\nonumber   \\
\frac {\partial u^Y}{\partial t} & = & v^Y .
\label{eq:LinHookeImplicit}
\end{eqnarray}
We use the same notation (\ref{eq:LinDiscreteDisp}) to represent the discretized displacements,
and to them we add the notation
\begin{equation}
V_{i,j}^k \equiv v^{1X} \left( x_i , y_j , t_k \right)
\quad \textrm{;} \quad
W_{i,j}^k \equiv v^{1Y} \left( x_i , y_j , t_k \right)
\label{eq:LinDiscreteVeloc}
\end{equation}
for discrete values of the two velocity components.

Adapting the approach of Qin \cite{Qin}, we take the first two equations in the
system (\ref{eq:LinHookeImplicit}), representing the $x$-component of the momentum equation, and
discretize them at the half-time step $t_{k+1/2}$ using the Crank-Nicolson scheme
\begin{eqnarray}
& & \frac {V_{i,j}^{k+1} - V_{i,j}^k}{\Delta t} 
- \frac {1}{2} \alpha^2 \left( \delta_x^2 X_{i,j}^{k+1} + \delta_x^2 X_{i,j}^k \right)   
\nonumber   \\
& & - \left( \alpha^2 - 1 \right) \delta_{xy}^2 Y_{i,j}^{k+1/2} 
- \frac {1}{2} \left( \delta_y^2 X_{i,j}^{k+1} + \delta_y^2 X_{i,j}^k \right) = 0 ,   
\nonumber   \\
& & \frac {1}{2} \left( V_{i,j}^{k+1} + V_{i,j}^k \right)
= \frac {X_{i,j}^{k+1} - X_{i,j}^k}{\Delta t}   .
\label{eq:LinImplicitXmom}
\end{eqnarray}
The second equation in this system (\ref{eq:LinImplicitXmom}) gives
\begin{equation}
X_{i,j}^{k+1} = X_{i,j}^k + \frac {1}{2} \Delta t \left( V_{i,j}^k
+ V_{i,j}^{k+1} \right)  ,
\label{eq:LinImplicitXmom2}
\end{equation}
from which the first in the system (\ref{eq:LinImplicitXmom}) can be written 
in operator notation as
\begin{eqnarray}
& & \left[ 1 - \frac {1}{4} \alpha^2 \Delta t^2 \delta_x^2 
- \frac {1}{4} \Delta t^2 \delta_y^2 \right] V_{i,j}^{k+1} 
\nonumber   \\
& & = \left[ 1 + \frac {1}{4} \alpha^2 \Delta t^2 \delta_x^2 
+ \frac {1}{4} \Delta t^2 \delta_y^2 \right] V_{i,j}^k 
\nonumber   \\
& & + \Delta t \left[ \alpha^2 \delta_x^2 X_{i,j}^k + \delta_y^2 X_{i,j}^k
+ \left( \alpha^2 - 1 \right) \delta_{xy}^2 Y_{i,j}^{k+ 1/2} \right]  .
\nonumber
\end{eqnarray}
Since this equation is accurate to second order in the grid spacings, it is
possible to add terms to each side that are of order $\Delta t^4$, without changing
the accuracy of the scheme.  By this means, we write an equation of equivalent
accuracy in factorized form
\begin{eqnarray}
& & \left[ 1- \frac {1}{4} \Delta t^2 \delta_y^2 \right] 
\left[ 1 - \frac {1}{4} \alpha^2 \Delta t^2 \delta_x^2 \right] V_{i,j}^{k+1} 
\nonumber   \\
& & = \left[ 1 + \frac {1}{4} \Delta t^2 \delta_y^2 \right] 
\left[ 1 + \frac {1}{4} \alpha^2 \Delta t^2 \delta_x^2 \right] V_{i,j}^k 
\nonumber   \\
& & + \Delta t \left[ \alpha^2 \delta_x^2 X_{i,j}^k + \delta_y^2 X_{i,j}^k
+ \left( \alpha^2 - 1 \right) \delta_{xy}^2 Y_{i,j}^{k+ 1/2} \right]  .
\label{eq:LinXmomCrank}
\end{eqnarray}
From this expression (\ref{eq:LinXmomCrank}), we derive an alternating-direction 
implicit (ADI) scheme, of Douglas type, as suggested by Qin \cite{Qin}.  There are 
two stages to the scheme at each new time step, and each stage involves a tri-diagonal
matrix equation.  The ADI method for the $x$-component of momentum consists of 
the first step
\begin{eqnarray}
& & \left[ 1 - \frac {1}{4} \Delta t^2 \delta_y^2 \right] \overline{V_{i,j}}
= \frac {1}{2} \Delta t^2 \left[ \alpha^2 \delta_x^2 + \delta_y^2 \right] V_{i,j}^k 
\nonumber   \\
& & + \Delta t \left[ \alpha^2 \delta_x^2 X_{i,j}^k + \delta_y^2 X_{i,j}^k
+ \left( \alpha^2 - 1 \right) \delta_{xy}^2 Y_{i,j}^{k+ 1/2} \right]  ,
\label{eq:LinXmomTriDiag1st}
\end{eqnarray}
followed by the second step
\begin{equation}
\left[ 1 - \frac {1}{4} \alpha^2 \Delta t^2 \delta_x^2 \right] \widehat{V_{i,j}}
= \overline{V_{i,j}}  .
\label{eq:LinXmomTriDiag2nd}
\end{equation}
In these equations, 
\begin{equation}
\overline{V_{i,j}} = V_{i,j}^{k+1/2} - V_{i,j}^k  \quad \textrm{and }  \quad
\widehat{V_{i,j}} = V_{i,j}^{k+1} - V_{i,j}^k  .
\label{eq:LinVoverbar}
\end{equation}

A similar process occurs for the $y$-component of the elastodynamic momentum
equation.  The second pair of partial differential equations in the 
system (\ref{eq:LinHookeImplicit}) is discretized at the half-time step $t_{k+1/2}$ 
using a Crank-Nicholson scheme, similar to (\ref{eq:LinImplicitXmom}), from which 
it follows that
\begin{equation}
Y_{i,j}^{k+1} = Y_{i,j}^k + \frac {1}{2} \Delta t \left( W_{i,j}^k
+ W_{i,j}^{k+1} \right)  ,
\label{eq:LinImplicitYmom2}
\end{equation}
as for equation (\ref{eq:LinImplicitXmom2}).  The discretized $y$-momentum equation 
can again be factorized, in operator form, by adding higher-order terms to each side
of the difference equation without affecting its accuracy.  This results in
\begin{eqnarray}
& & \left[ 1- \frac {1}{4} \alpha^2 \Delta t^2 \delta_y^2 \right] 
\left[ 1 - \frac {1}{4} \Delta t^2 \delta_x^2 \right] W_{i,j}^{k+1} 
\nonumber   \\
& & = \left[ 1 + \frac {1}{4} \alpha^2 \Delta t^2 \delta_y^2 \right] 
\left[ 1 + \frac {1}{4} \Delta t^2 \delta_x^2 \right] W_{i,j}^k 
\nonumber   \\
& & + \Delta t \left[ \alpha^2 \delta_y^2 Y_{i,j}^k + \delta_x^2 Y_{i,j}^k
+ \left( \alpha^2 - 1 \right) \delta_{xy}^2 X_{i,j}^{k+ 1/2} \right]  ,
\label{eq:LinYmomCrank}
\end{eqnarray}
similar to (\ref{eq:LinXmomCrank}).  Operator splitting is again invoked, to produce a
two-stage ADI scheme for the $y$-component of momentum, of the form
\begin{eqnarray}
& & \left[ 1 - \frac {1}{4} \alpha^2 \Delta t^2 \delta_y^2 \right] \overline{W_{i,j}}
= \frac {1}{2} \Delta t^2 \left[ \delta_x^2 + \alpha^2 \delta_y^2 \right] W_{i,j}^k 
\nonumber   \\
& & + \Delta t \left[ \alpha^2 \delta_y^2 Y_{i,j}^k + \delta_x^2 Y_{i,j}^k
+ \left( \alpha^2 - 1 \right) \delta_{xy}^2 X_{i,j}^{k+ 1/2} \right]  ,
\label{eq:LinYmomTriDiag1st}
\end{eqnarray}
followed by
\begin{equation}
\left[ 1 - \frac {1}{4} \Delta t^2 \delta_x^2 \right] \widehat{W_{i,j}}
= \overline{W_{i,j}}  ,
\label{eq:LinYmomTriDiag2nd}
\end{equation}
in which we have similarly defined intermediate quantities 
\begin{equation}
\overline{W_{i,j}} = W_{i,j}^{k+1/2} - W_{i,j}^k  \quad \textrm{and }  \quad
\widehat{W_{i,j}} = W_{i,j}^{k+1} - W_{i,j}^k  ,
\label{eq:LinWoverbar}
\end{equation}
as in equation (\ref{eq:LinVoverbar}).

The calculation of variables at each new time step therefore involves
first computing and storing intermediate quantities $\overline{V_{i,j}}$ and
$\overline{W_{i,j}}$ from (\ref{eq:LinXmomTriDiag1st}) and (\ref{eq:LinYmomTriDiag1st}).  
The right-hand sides of these equations require displacements at the half-time step, and 
these are first estimated here by extrapolation, using the previous two time steps.  Thus
\begin{eqnarray}
X_{i,j}^{k+ 1/2} & = & \frac {3}{2} X_{i,j}^k - \frac {1}{2} X_{i,j}^{k-1} 
\nonumber   \\
Y_{i,j}^{k+ 1/2} & = & \frac {3}{2} Y_{i,j}^k - \frac {1}{2} Y_{i,j}^{k-1} .
\label{eq:LinXYextrap}
\end{eqnarray}
While these estimates (\ref{eq:LinXYextrap}) admittedly do not conform to 
Qin's \cite{Qin} concept of a ``compact'' ADI scheme, they nevertheless provide the 
required information at the necessary degree of accuracy.  Once these
intermediate quantities $\overline{V_{i,j}}$ and $\overline{W_{i,j}}$
have thus been computed, the further two tri-diagonal systems
(\ref{eq:LinXmomTriDiag2nd}) and (\ref{eq:LinYmomTriDiag2nd}) are then solved 
to give $\widehat{V_{i,j}}$ and $\widehat{W_{i,j}}$, and then 
equations (\ref{eq:LinVoverbar}), (\ref{eq:LinWoverbar}) at once yield
$V_{i,j}^{k+1}$ and $W_{i,j}^{k+1}$ at the new time step.  Finally,
the two displacements $X_{i,j}^{k+1}$ and $Y_{i,j}^{k+1}$ are
calculated at this new time using (\ref{eq:LinImplicitXmom2}) 
and (\ref{eq:LinImplicitYmom2}).  We observe that the ``false boundary'' 
$y_{N+1} = \Delta y$ above the surface is again involved in these implicit equations, 
and displacements there are obtained from the two dynamic boundary 
conditions (\ref{eq:LinTangentExplicit}), (\ref{eq:LinNormalExplicit}) 
expressed here as 
\begin{eqnarray}
X_{i,N+1}^{k+1} & = & X_{i,N-1}^{k+1} - 2 \Delta y \delta_x Y_{i,N}^{k+1} 
\nonumber   \\
Y_{i,N+1}^{k+1} & = & Y_{i,N-1}^{k+1} 
- 2 \Delta y \left( \frac {\alpha^2 - 2}{\alpha^2} \right) \delta_x X_{i,N}^{k+1} ,
\label{eq:LinXYfalseboundary}
\end{eqnarray}
in which it is convenient here, and in later work, to introduce the first-order 
difference operators
\begin{eqnarray}
\delta_x X_{i,j}^k 
& = & \left( X_{i+1,j}^k - X_{i-1,j}^k \right) / \left( 2\Delta x \right) 
\nonumber   \\
\delta_y X_{i,j}^k 
& = & \left( X_{i,j+1}^k - X_{i,j-1}^k \right) / \left( 2\Delta y \right) .
\label{eq:LinDXDYoperator}
\end{eqnarray}
The two velocity components $V_{i,N+1}^{k+1}$ and $W_{i,N+1}^{k+1}$ at
this false boundary are obtained by re-arranging equations (\ref{eq:LinImplicitXmom2}), 
(\ref{eq:LinImplicitYmom2}) with $j$ set to $N+1$.

\section{The Nonlinear System}
\label{sec:nonlinear}

The nonlinear problem is given by equations (\ref{eq:HookeComponent}), (\ref{eq:Kinem}), 
(\ref{eq:DynamicNormal})--(\ref{eq:InitialVeloc}), and the principal difficulty 
is that it involves a solution domain $y < S \left( x,t \right)$ that is 
unknown {\it a priori}.  Closed-form mathematical solutions are not feasible, 
and so to enable a numerical approach, we transform from coordinates $\left( x,y \right)$ 
to the non-orthogonal system $\left( x , \eta \right)$, in which 
\begin{equation}
\eta \left( x,y,t \right) = y - S \left( x,t \right)   .
\label{eq:NonDefineEta}
\end{equation}
Now the free surface corresponds to the curve $\eta = 0$.  In these new
coordinates, the $x$-Momentum equation in (\ref{eq:HookeComponent}) becomes
\begin{eqnarray}
& & \frac {\partial v^X}{\partial t} 
- \frac{\partial u^X}{\partial \eta} \left( \frac {\partial^2 S}{\partial t^2} \right)
- 2 \frac{\partial v^X}{\partial \eta} \left( \frac {\partial S}{\partial t} \right)
+ \frac{\partial^2 u^X}{\partial \eta^2} \left( \frac {\partial S}{\partial t} \right)^2 
\nonumber   \\
& = & \alpha^2 \biggl[ \frac {\partial^2 u^X}{\partial x^2} 
- 2 \frac{\partial^2 u^X}{\partial x \partial \eta} \left( \frac {\partial S}{\partial x} \right)
\nonumber   \\
& &  + \frac{\partial^2 u^X}{\partial \eta^2} \left( \frac {\partial S}{\partial x} \right)^2
- \frac{\partial u^X}{\partial \eta} \left( \frac {\partial^2 S}{\partial x^2} \right)
\biggr]  
\nonumber   \\
& + & \left( \alpha^2 - 1 \right) \left[ \frac {\partial^2 u^Y}{\partial x \partial \eta}
- \frac{\partial^2 u^Y}{\partial \eta^2} \left( \frac {\partial S}{\partial x} \right)
\right]
+ \frac {\partial^2 u^X}{\partial \eta^2} .
\label{eq:NonLinXmom}
\end{eqnarray}
The $y$-Momentum equation in (\ref{eq:HookeComponent}) now becomes
\begin{eqnarray}
& & \frac {\partial v^Y}{\partial t} 
- \frac{\partial u^Y}{\partial \eta} \left( \frac {\partial^2 S}{\partial t^2} \right)
- 2 \frac{\partial v^Y}{\partial \eta} \left( \frac {\partial S}{\partial t} \right)
+ \frac{\partial^2 u^Y}{\partial \eta^2} \left( \frac {\partial S}{\partial t} \right)^2 
\nonumber   \\
& = & \alpha^2 \frac {\partial^2 u^Y}{\partial \eta^2}
+ \left( \alpha^2 - 1 \right) \left[ \frac {\partial^2 u^X}{\partial x \partial \eta}
- \frac{\partial^2 u^X}{\partial \eta^2} \left( \frac {\partial S}{\partial x} \right)
\right]  
\nonumber   \\
& + & \frac {\partial^2 u^Y}{\partial x^2} 
- 2 \frac{\partial^2 u^Y}{\partial x \partial \eta} \left( \frac {\partial S}{\partial x} \right)
+ \frac{\partial^2 u^Y}{\partial \eta^2} \left( \frac {\partial S}{\partial x} \right)^2
\nonumber   \\
& - & \frac{\partial u^Y}{\partial \eta} \left( \frac {\partial^2 S}{\partial x^2} \right)  .
\label{eq:NonLinYmom}
\end{eqnarray}
In these expressions (\ref{eq:NonLinXmom}), (\ref{eq:NonLinYmom}), we have defined 
auxiliary variables
\begin{equation}
v^X = \frac {\partial u^X}{\partial t}  \quad \textrm{;} \quad
v^Y = \frac {\partial u^Y}{\partial t}
\label{eq:NonLinDefineVXVY}
\end{equation}
similar to those in (\ref{eq:LinHookeImplicit}), although here the time derivatives 
hold $\eta$ constant, rather than $y$, so that these quantities (\ref{eq:NonLinDefineVXVY}) 
are not simply cartesian velocity components.

The kinematic boundary condition (\ref{eq:Kinem}) transforms to
\begin{equation}
\frac {\partial S}{\partial t} 
= \frac {\left[ \left( \partial u^Y / \partial t \right)
- \left( \partial u^X / \partial t \right) \left( \partial S / \partial x \right) \right]}
{\left[ 1 + \left( \partial u^Y / \partial \eta \right)
- \left( \partial u^X / \partial \eta \right) \left( \partial S / \partial x \right) \right]}
\quad \textrm{on } \eta = 0  .
\label{eq:NonLinKinem}
\end{equation}
The normal dynamic condition (\ref{eq:DynamicNormal}) at the free surface becomes
\begin{eqnarray}
& & \left( \frac {\partial S}{\partial x} \right)^2 \left[ 
\alpha^2 \frac {\partial u^X}{\partial x}
- \alpha^2 \frac {\partial u^X}{\partial \eta} \frac {\partial S}{\partial x}
+ \left( \alpha^2 - 2 \right) \frac {\partial u^Y}{\partial \eta} \right] 
\nonumber   \\
& - & 2 \frac {\partial S}{\partial x} \left[ \frac {\partial u^X}{\partial \eta} 
+ \frac {\partial u^Y}{\partial x}
- \frac {\partial u^Y}{\partial \eta} \frac{\partial S}{\partial x} \right] 
\nonumber   \\
& + & \left( \alpha^2 - 2 \right) \left[ \frac {\partial u^X}{\partial x}
- \frac {\partial u^X}{\partial \eta} \frac {\partial S}{\partial x} \right]
+ \alpha^2 \frac {\partial u^Y}{\partial \eta} = 0
\nonumber   \\
& &  \quad \textrm{on } \eta = 0  ,
\label{eq:NonLinDynNormal}
\end{eqnarray}
and the tangential dynamic free surface condition (\ref{eq:DynamicTangent}) takes the form
\begin{eqnarray}
& & 2 \frac {\partial S}{\partial x} \left[ \frac {\partial u^Y}{\partial \eta} 
- \frac {\partial u^X}{\partial x} 
+ \frac {\partial u^X}{\partial \eta} \frac {\partial S}{\partial x} \right]  
\nonumber   \\
& + & \left( 1 - \left( \frac {\partial S}{\partial x} \right)^2 \right) \left[ 
\frac {\partial u^X}{\partial \eta} + \frac {\partial u^Y}{\partial x} 
- \frac {\partial u^Y}{\partial \eta} \frac {\partial S}{\partial x} \right] = 0
\nonumber   \\
& &  \quad \textrm{on } \eta = 0  .
\label{eq:NonLinDynTangent}
\end{eqnarray}
These two dynamic conditions (\ref{eq:NonLinDynNormal}) and (\ref{eq:NonLinDynTangent}) at 
the surface are combined and re-arranged to give the normal derivatives
\begin{eqnarray}
\frac {\partial u^X}{\partial \eta} & = &
\frac {\partial u^X}{\partial x} \frac {S_x \left[ \alpha^2 S_x^2 + 3\alpha^2 -2 \right]}
{\alpha^2 \left[ 1 + S_x^2 \right]^2} 
\nonumber   \\
 & + & \frac {\partial u^Y}{\partial x} 
\frac {\left[ \left( \alpha^2 - 2 \right) S_x^2 - \alpha^2 \right]}
{\alpha^2 \left[ 1 + S_x^2 \right]^2}
\nonumber   \\
& &   \quad \textrm{on } \eta = 0  ,
\label{eq:NonLinDuxDeta}
\end{eqnarray}
and
\begin{equation}
\frac {\partial u^Y}{\partial \eta} = \left[
\frac {\partial u^X}{\partial x} + S_x \frac {\partial u^Y}{\partial x} \right]
\frac {\left[ \alpha^2 S_x^2 - \alpha^2 + 2 \right]}{\alpha^2 \left[ 1 + S_x^2 \right]^2}
\quad \textrm{on } \eta = 0  .
\label{eq:NonLinDuyDeta}
\end{equation}
In these expressions, the subscripts again indicate partial derivatives.
These new forms (\ref{eq:NonLinDuxDeta}), (\ref{eq:NonLinDuyDeta}) of the dynamic 
boundary conditions are substituted into the denominator in (\ref{eq:NonLinKinem}), and 
after some algebra, the kinematic condition is obtained as
\begin{eqnarray}
& & \frac {\partial S}{\partial t} =
\nonumber   \\
& & \frac {\alpha^2 \left[ 1 + S_x^2 \right] \left[ \left( \partial u^Y / \partial t \right)
- S_x \left( \partial u^X / \partial t \right) \right]}
{\alpha^2 \left[ 1 + S_x^2 \right] \left[ 1 - \left( \partial u^X / \partial x \right) \right]
+ 2 \left[ \left( \partial u^X / \partial x \right) 
+ S_x \left( \partial u^Y / \partial x \right) \right]}
\nonumber   \\
& &    \quad \textrm{on } \eta = 0  .
\label{eq:NonLinKinemNew}
\end{eqnarray}
This form (\ref{eq:NonLinKinemNew}) is more amenable to numerical treatment 
than (\ref{eq:NonLinKinem}).

To solve this highly nonlinear problem, we have adapted the ADI method described in 
Section \ref{sec:implicit}, and combined it with a predictor--corrector type scheme 
to account for the nonlinear terms at the surface.  To integrate from time level $t = t_k$
to the next level $t = t_{k+1}$ we first estimate solution variables at the half-time
$t_{k+1/2}$ using extrapolation, as in (\ref{eq:LinXYextrap}), and to these we also add
\begin{equation}
S_i^{k+1/2} =  \frac {3}{2} S_i^k - \frac {1}{2} S_i^{k-1}
\nonumber
\end{equation}
for the free-surface elevation.  We then calculate and store intermediate functions
\begin{eqnarray}
M_i^k & = & \frac {1}{4} \left[ \left( S_{tt} \right)_i^{k+1/2}
- \alpha^2 \left( S_{xx} \right)_i^{k+1/2} \right]
\nonumber   \\
N_i^k & = & \frac {1}{4} \left[ 1 + \alpha^2 \left( S_x^2 \right)_i^{k+1/2}
- \left( S_t^2 \right)_i^{k+1/2} \right] 
\nonumber   \\
P_i^k & = & \left( S_{t} \right)_i^{k+1/2}
\nonumber   \\
R_i^k & = & \frac {1}{4} \left[ \left( S_{tt} \right)_i^{k+1/2}
- \left( S_{xx} \right)_i^{k+1/2} \right] 
\nonumber   \\
T_i^k & = & \frac {1}{4} \left[ \alpha^2 + \left( S_x^2 \right)_i^{k+1/2}
- \left( S_t^2 \right)_i^{k+1/2} \right]   .
\label{eq:NonLinInterVariables}
\end{eqnarray}
Following the similar procedure used for the linearized equations, the ADI scheme
for the $x$-momentum equation (\ref{eq:NonLinXmom}) takes the two-stage form
\begin{eqnarray}
& & \left[ 1 - \Delta t P_i^k \delta_{\eta} - \Delta t^2 M_i^k \delta_{\eta}
- \Delta t^2 N_i^k \delta_{\eta}^2 \right] \left[ V_{i,j}^{k+1/2} - V_{i,j}^k \right] 
\nonumber   \\
& = & 2 \left[ \Delta t P_i^k \delta_{\eta} + \Delta t^2 M_i^k \delta_{\eta}
+ \Delta t^2 N_i^k \delta_{\eta}^2 + \frac {1}{4} \alpha^2 \Delta t^2 \delta_x^2
\right] V_{i,j}^k  
\nonumber   \\
& + & 4 \Delta t \left[ M_i^k \delta_{\eta} X_{i,j}^k 
+ N_i^k \delta_{\eta}^2 X_{i,j}^k \right] + \alpha^2 \Delta t \delta_x^2 X_{i,j}^k 
\nonumber   \\
& - & \Delta t \left( S_x \right)_i^{k+1/2} \left[ 
2\alpha^2 \delta_{x \eta}^2 X_{i,j}^{k+1/2} 
+ \left( \alpha^2 -1 \right) \delta_{\eta}^2 Y_{i,j}^{k+1/2} \right]
\nonumber   \\
& + & \left( \alpha^2 -1 \right) \Delta t \delta_{x \eta}^2 Y_{i,j}^{k+1/2}  ,
\label{eq:NonLinXmomTriDiag1st}
\end{eqnarray}
which is a tri-diagonal system to be solved for 
$\overline{V_{i,j}} = V_{i,j}^{k+1/2} - V_{i,j}^k$ , followed by
\begin{equation}
\left[ 1 - \frac {1}{4} \alpha^2 \Delta t^2 \delta_x^2 \right]
\left[ V_{i,j}^{k+1} - V_{i,j}^k \right] = \overline{V_{i,j}}   .
\label{eq:NonLinXmomTriDiag2nd}
\end{equation}
This is also a tri-diagonal system for $\widehat{V_{i,j}} = V_{i,j}^{k+1} - V_{i,j}^k$.
We observe that the two differential operators in these 
expressions (\ref{eq:NonLinXmomTriDiag1st}) and (\ref{eq:NonLinXmomTriDiag2nd}) 
do not commute, and so the order in which these two systems of equations
is written and solved is important.

For completeness, we also give here the equivalent ADI scheme for the 
$y$-Momentum equation (\ref{eq:NonLinYmom}).  The first system is
\begin{eqnarray}
& & \left[ 1 - \Delta t P_i^k \delta_{\eta} - \Delta t^2 R_i^k \delta_{\eta}
- \Delta t^2 T_i^k \delta_{\eta}^2 \right] \left[ W_{i,j}^{k+1/2} - W_{i,j}^k \right] 
\nonumber   \\
& = & 2 \left[ \Delta t P_i^k \delta_{\eta} + \Delta t^2 R_i^k \delta_{\eta}
+ \Delta t^2 T_i^k \delta_{\eta}^2 + \frac {1}{4} \Delta t^2 \delta_x^2
\right] W_{i,j}^k  
\nonumber   \\
& + & 4 \Delta t \left[ R_i^k \delta_{\eta} Y_{i,j}^k 
+ T_i^k \delta_{\eta}^2 Y_{i,j}^k \right] + \Delta t \delta_x^2 Y_{i,j}^k 
\nonumber   \\
& - & \Delta t \left( S_x \right)_i^{k+1/2} \left[ 
2 \delta_{x \eta}^2 Y_{i,j}^{k+1/2} 
+ \left( \alpha^2 -1 \right) \delta_{\eta}^2 X_{i,j}^{k+1/2} \right]
\nonumber   \\
& + & \left( \alpha^2 -1 \right) \Delta t \delta_{x \eta}^2 X_{i,j}^{k+1/2}  ,
\label{eq:NonLinYmomTriDiag1st}
\end{eqnarray}
followed by
\begin{equation}
\left[ 1 - \frac {1}{4} \Delta t^2 \delta_x^2 \right]
\left[ W_{i,j}^{k+1} - W_{i,j}^k \right] = W_{i,j}^{k+1/2} - W_{i,j}^k   .
\label{eq:NonLinYmomTriDiag2nd}
\end{equation}
Once $V_{i,j}^{k+1}$ and $W_{i,j}^{k+1}$ have thus been found, the quantities
$X_{i,j}^{k+1}$ and $Y_{i,j}^{k+1}$ are determined from (\ref{eq:LinImplicitXmom2}) 
and (\ref{eq:LinImplicitYmom2}).  Once again, a ``false boundary'' 
at $\eta_{N+1} = \Delta \eta$ is needed above the free surface, and 
values on that curve are obtained using the normal and tangential 
boundary conditions (\ref{eq:NonLinDuxDeta}) and (\ref{eq:NonLinDuyDeta}) in
finite-difference form.  Finally, the predictor value for the surface
elevation is corrected, using the kinematic free-surface condition
(\ref{eq:NonLinKinemNew}), discretized (using the Crank-Nicholson approach) to yield
\begin{eqnarray}
S_i^{k+1} & = & S_i^k + \Delta t \biggl[ 
\frac {1}{2} \left( W_{i,N}^{k+1} + W_{i,N}^k \right) \Phi_i^k
\nonumber   \\
& & - \frac {1}{2} \left( V_{i,N}^{k+1} + V_{i,N}^k \right) \Psi_i^k \biggr]  ,
\label{eq:NonLinSurfCorrect}
\end{eqnarray}
where
\begin{eqnarray}
\Phi_i^k & = & 
\frac {\alpha^2 \left[ 1 + \left( \delta_x S_i^{k+1/2} \right)^2 \right]}{D_i^k}
\nonumber   \\
\Psi_i^k & = & \Phi_i^k \left( \delta_x S_i^{k+1/2} \right)  ,
\nonumber
\end{eqnarray}
and
\begin{eqnarray}
D_i^k & = & \alpha^2 \left[ 1 + \left( \delta_x S_i^{k+1/2} \right)^2 \right] \left[
1 - \delta_x X_{i,N}^{k+1/2} \right]
\nonumber   \\
& + & 2 \left[ \delta_x X_{i,N}^{k+1/2} 
+ \left( \delta_x S_i^{k+1/2} \right) \left( \delta_x Y_{i,N}^{k+1/2} \right) \right] .
\nonumber
\end{eqnarray}
These routines have been coded in the {\it MATLAB\/} environment, where they can
be run fairly routinely with many hundreds of mesh points in both the $x$ and 
$y$ coordinates.

\section{Results}
\label{sec:results}

The routines developed for the linearized problem in Section \ref{sec:linear} and
the non-linear problem in Section \ref{sec:nonlinear} have been programmed in the
{\it MATLAB\/} environment.  The explicit and implicit ADI methods in sections
\ref{sec:explicit} and \ref{sec:implicit}, for the linearized problem, agree
very closely for smaller values of $\alpha^2$, such as $\alpha^2 = 3$, but 
if $\alpha^2$ is increased to $\alpha^2 = 5$ then the explicit method becomes 
numerically unstable.  The implicit method in Section \ref{sec:implicit}
remains stable, however.  The initial disturbance is centred on the $y$-axis
a distance $1$ below the initial surface $y = 0$, and for definiteness we have
set its half-lengths in equation (\ref{eq:InitialVeloc}) to be $a = 0.25$ and $b = 0.25$.

\begin{figure}
\centering
\includegraphics[width=0.75\textwidth]{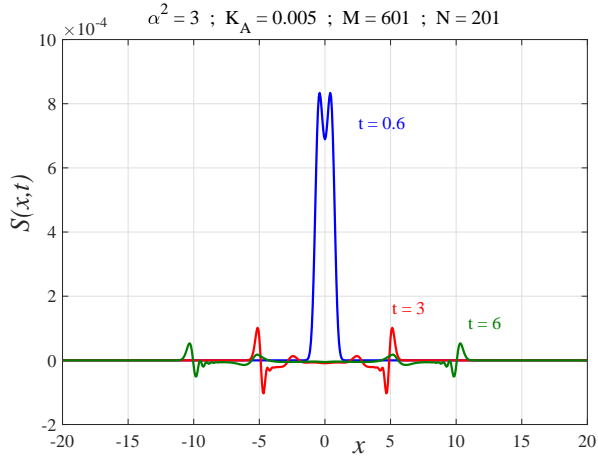}
\caption{Nonlinear free-surface elevation for $\alpha^2 = 3$, and
initial disturbance amplitude $K_A = 0.005$.  Surface waves are shown for three
times $t = 0.6$, $3$ and $6$.}
\label{fig_Figure1}
\end{figure}

Figure \ref{fig_Figure1} shows the nonlinear solution for speed ratio parameter
$\alpha^2 = 3$ (defined in (\ref{eq:ParamA})).  This solution was obtained over the 
computational domain $-20 < x < 20$, $-10 < \eta < 0$ with $M = 601$ points  in $x$
and $201$ points in $\eta$, with time steps $\Delta t = 0.004$.  The amplitude 
of the initial disturbance is $K_A = 0.005$ and free-surface elevations are shown 
for the three different times $t = 0.6$, $3$ and $6$.   Each solution 
is left-right symmetric, as is expected.  Initially, the surface is the line $y = 0$ 
and as time progresses, a single elevation develops above the location of the
initial disturbance, with its maximum at $x = 0$.  A little later, a dip forms
near the crest at $x = 0$ and it continues to develop at later times.  This dip
is visible at the earliest time $t = 0.6$ in Figure \ref{fig_Figure1}.
After some time, this central peak divides into two separate waves, one moving
to the right and the other to the left.  These two separate waves are visible at
(dimensionless) time $t = 3$ in Figure \ref{fig_Figure1}.  As these
waves continue to move away to the left and the right, their amplitudes
continue to decrease, and this is evident for the profile at the last time
$t = 6$ in Figure \ref{fig_Figure1}.

\begin{figure}
\centering
\includegraphics[width=0.6\textwidth]{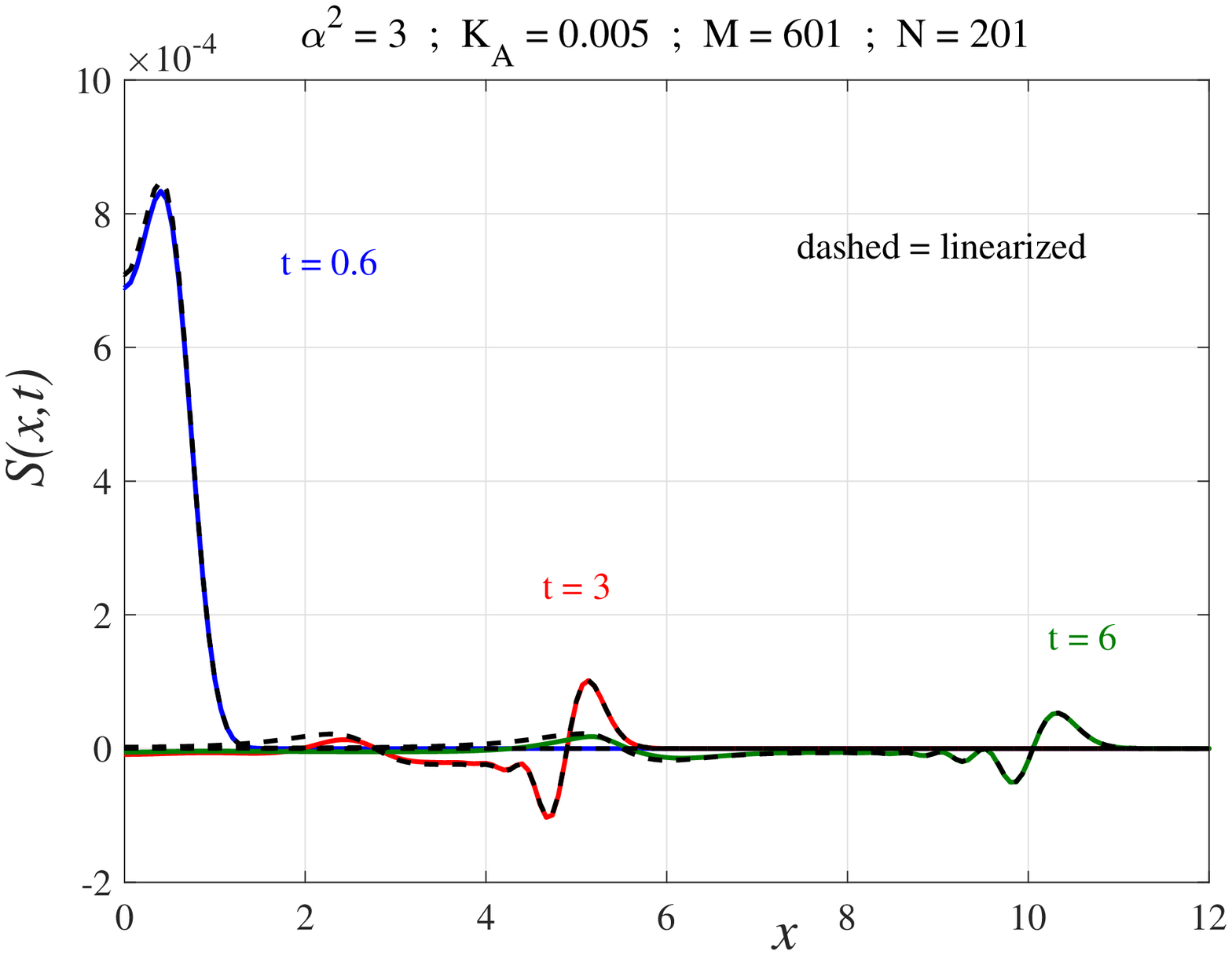}
\includegraphics[width=0.6\textwidth]{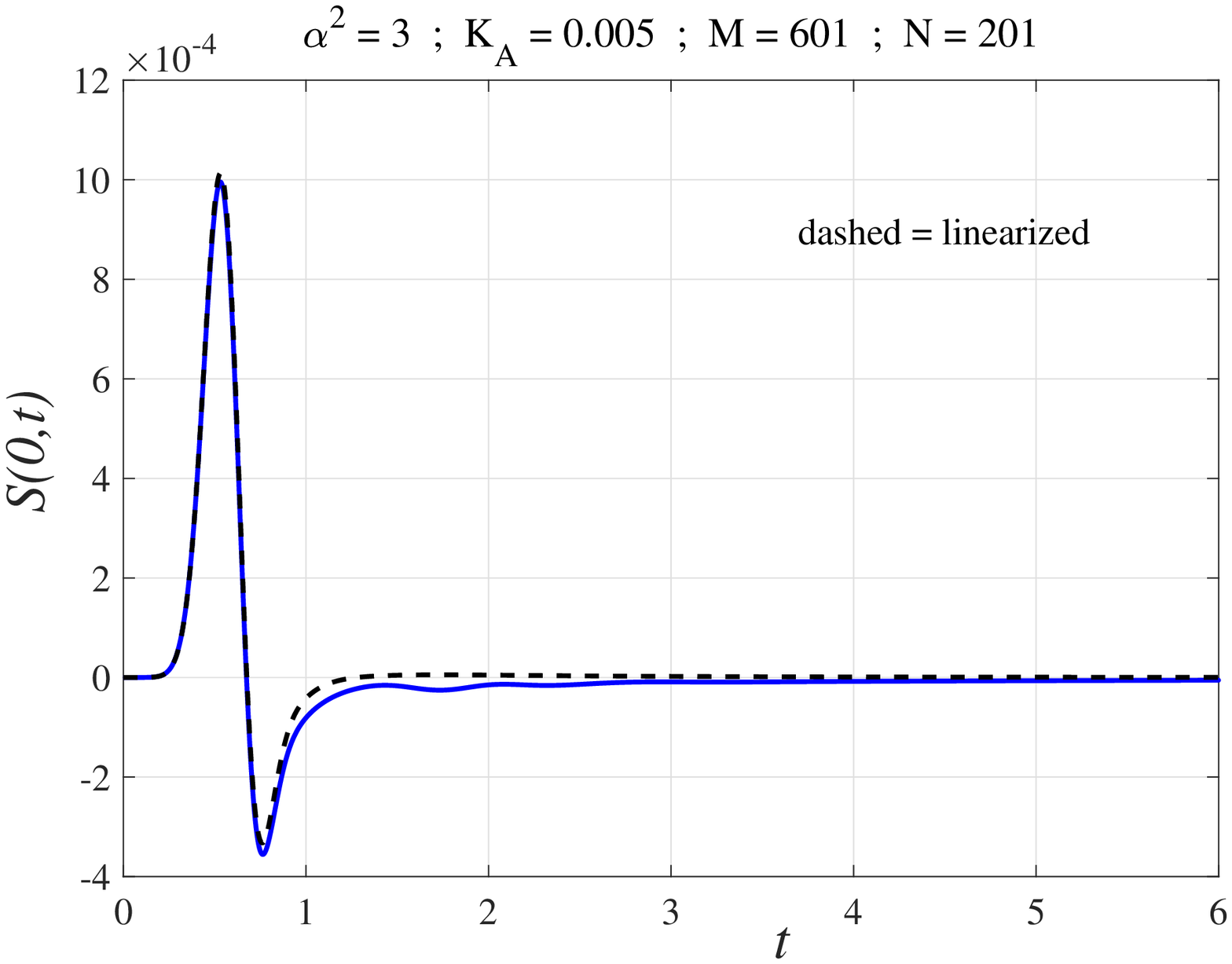}
\caption{A comparison of the linearized and nonlinear solutions for
the case $\alpha^2 = 3$, $K_A = 0.005$, as in Figure \ref{fig_Figure1}.  
(a) Surface profiles are shown for the three times $t = 0.6$, $3$ and $6$, and 
the linearized elevations are sketched with dashed lines;  
(b) The time history of the surface elevation $S$ at the centre point $x = 0$.}
\label{fig_Figure2}
\end{figure}

It is important to verify the numerical reliability of the nonlinear
ADI algorithm of Section \ref{sec:nonlinear}.  For the small initial amplitude 
$K_A = 0.005$ in Figure \ref{fig_Figure1}, the linearized and nonlinear
results ought to be very similar, and this is shown in Figure \ref{fig_Figure2}.
In part (a), the three nonlinear solutions are shown exactly as in 
Figure \ref{fig_Figure1} for the same three times, although here only the 
right-hand side is displayed by symmetry, to permit a clearer view of the 
wave elevations.  The linearized solution is also shown for these three times, 
and is sketched with dashed lines.  We find that the linearized solution slightly 
over-estimates the height of the single peak that first forms over the disturbance 
at $x = 0$, and this effect becomes stronger as the initial amplitude $K_A$ is
increased.  But as the central peak decreases in amplitude, splits and 
then spawns a left-moving and right-moving wave, the agreement with the 
linearized solution becomes better and particularly so at the outward-moving 
fronts.  This is clearly evident in Figure \ref{fig_Figure2}(a).  This behaviour
is studied in more detail in part (b), for the surface elevation at the centre
point $x = 0$.  Here, the function $S \left( 0,t \right)$ has been recorded at
every time interval calculated, and is shown as a function of time $t$.  The
linearized solution is again drawn with a dashed line.  There is close agreement
between the two for all times except for a window at about $t = 2$, where the
nonlinear solution develops a small secondary wave that is absent in the
linearized case.

\begin{figure}
\centering
\includegraphics[width=0.6\textwidth]{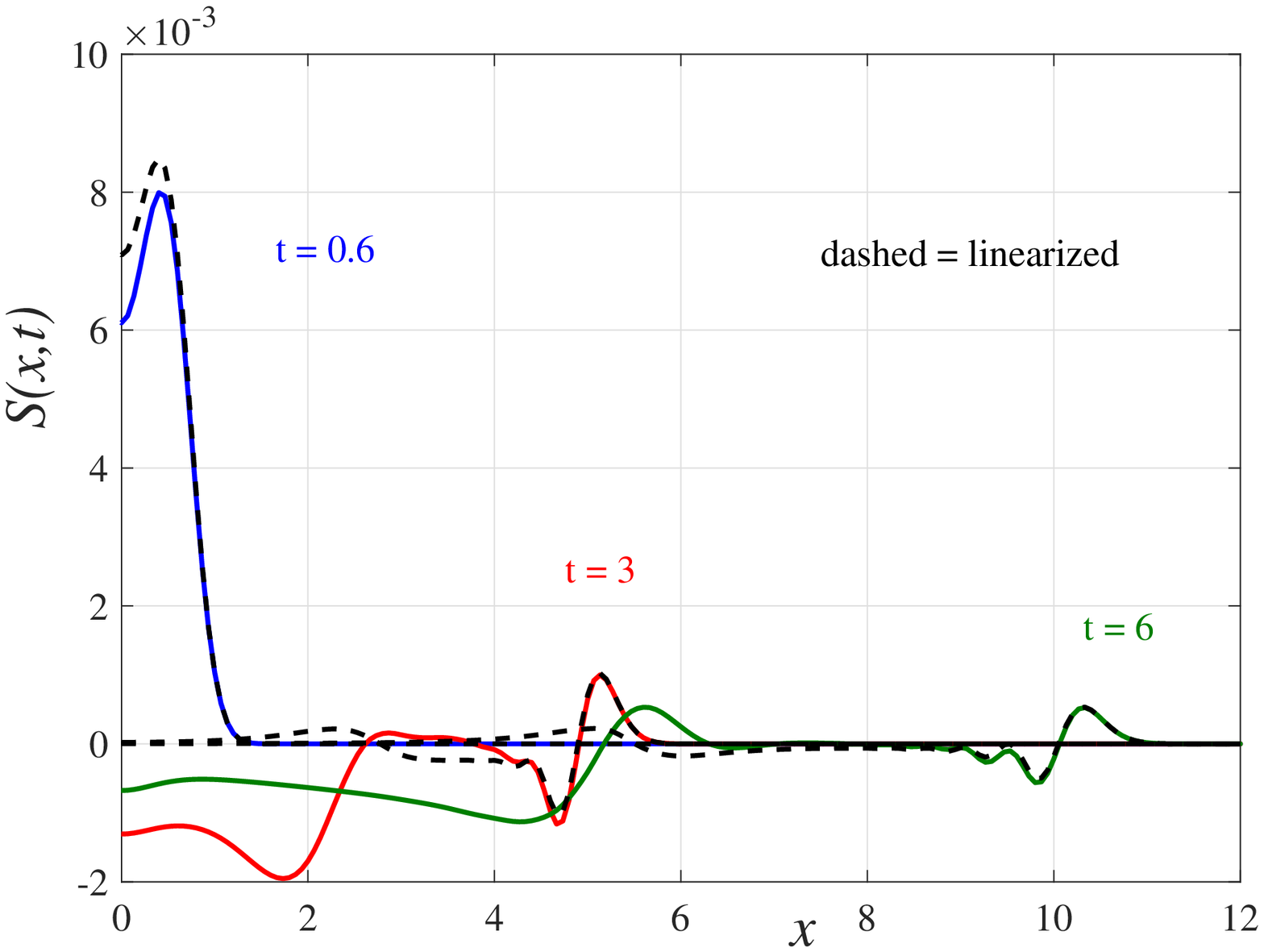}
\includegraphics[width=0.6\textwidth]{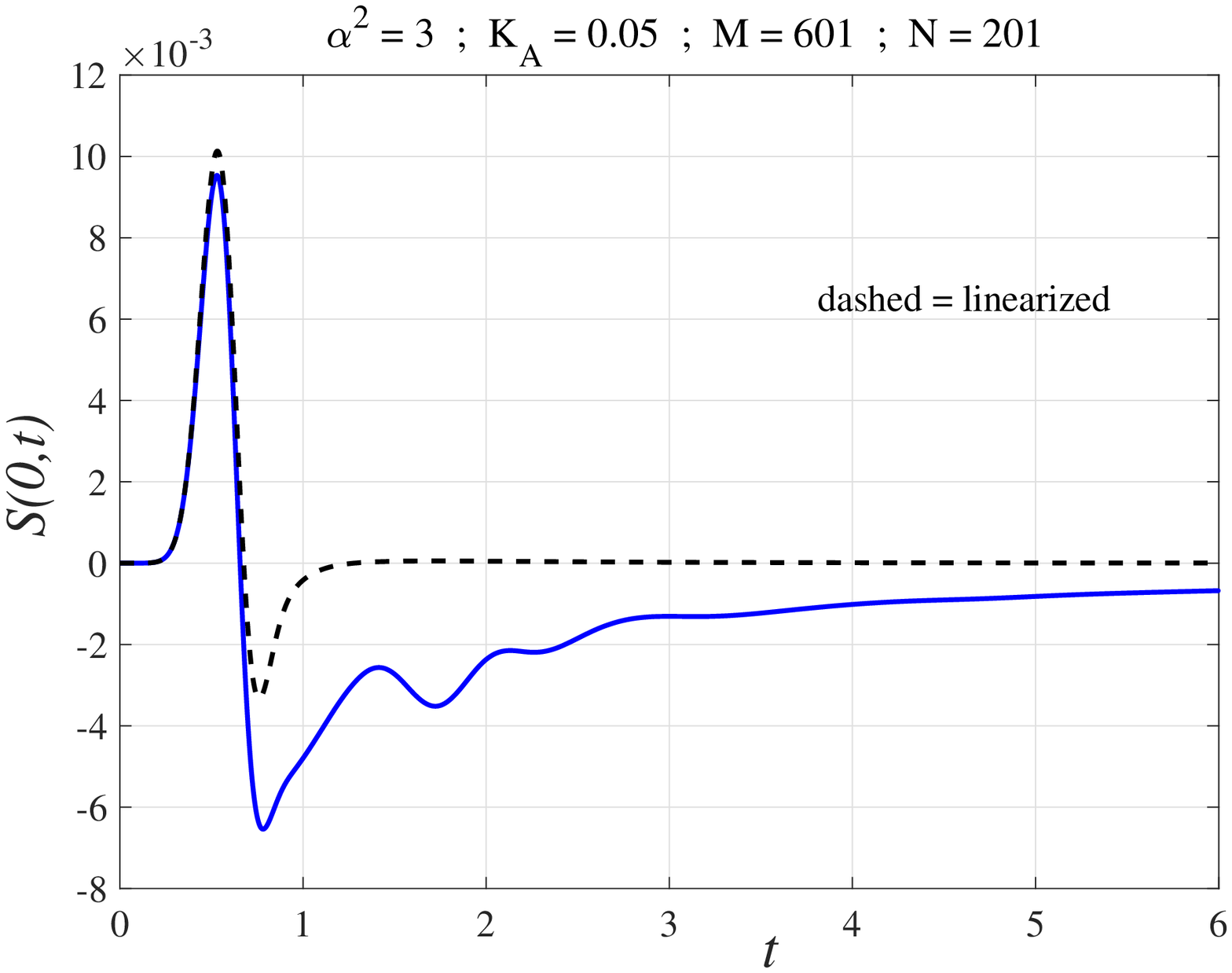}
\caption{A comparison of the linearized and nonlinear solutions for
speed ratio $\alpha^2 = 3$ and initial disturbance amplitude $K_A = 0.05$.  
(a) Surface profiles are shown for the three times $t = 0.6$, $3$ and $6$, and 
the linearized elevations are sketched with dashed lines;
(b) The time history of the surface elevation $S$ at the centre point $x = 0$.}
\label{fig_Figure3}
\end{figure}

A solution at the same speed ratio $\alpha^2 = 3$, but much greater
initial disturbance amplitude $K_A = 0.05$, is displayed in Figure 
\ref{fig_Figure3}.  The behaviour of the free surface in part (a) is similar to 
that at the much smaller amplitude in previous Figure \ref{fig_Figure2}(a).
Initially the surface is flat, a single hump then grows above the location 
of the disturbance, and the hump then splits into two separate waves that
propagate in opposite directions while slowly losing amplitude.  At the
earliest time $t = 0.6$ shown in Figure \ref{fig_Figure3}(a) there is again
good agreement between the linearized and nonlinear solutions, except that
the linearized solution over-estimates the height of the initial hump.
However, as the waves propagate outwards at later times, the agreement
between the two solutions near the outward-moving wave fronts is very close.
The striking difference between the two solutions, however, is that the
linearized solution, sketched with dashed lines, very substantially 
under-estimates the extent of the rebound of the free surface near $x = 0$,
in the region above the disturbance.  This is studied in more detail in 
part (b), where it can be seen that the rebound is strongest at about
time $t = 0.8$.  Both the linearized and nonlinear surfaces return toward
zero at later times, although now the nonlinear solution shows several smaller
wavelets at $x = 0$, in the approximate time interval  $1.5 < t < 3$.

\begin{figure}
\centering
\includegraphics[width=0.6\textwidth]{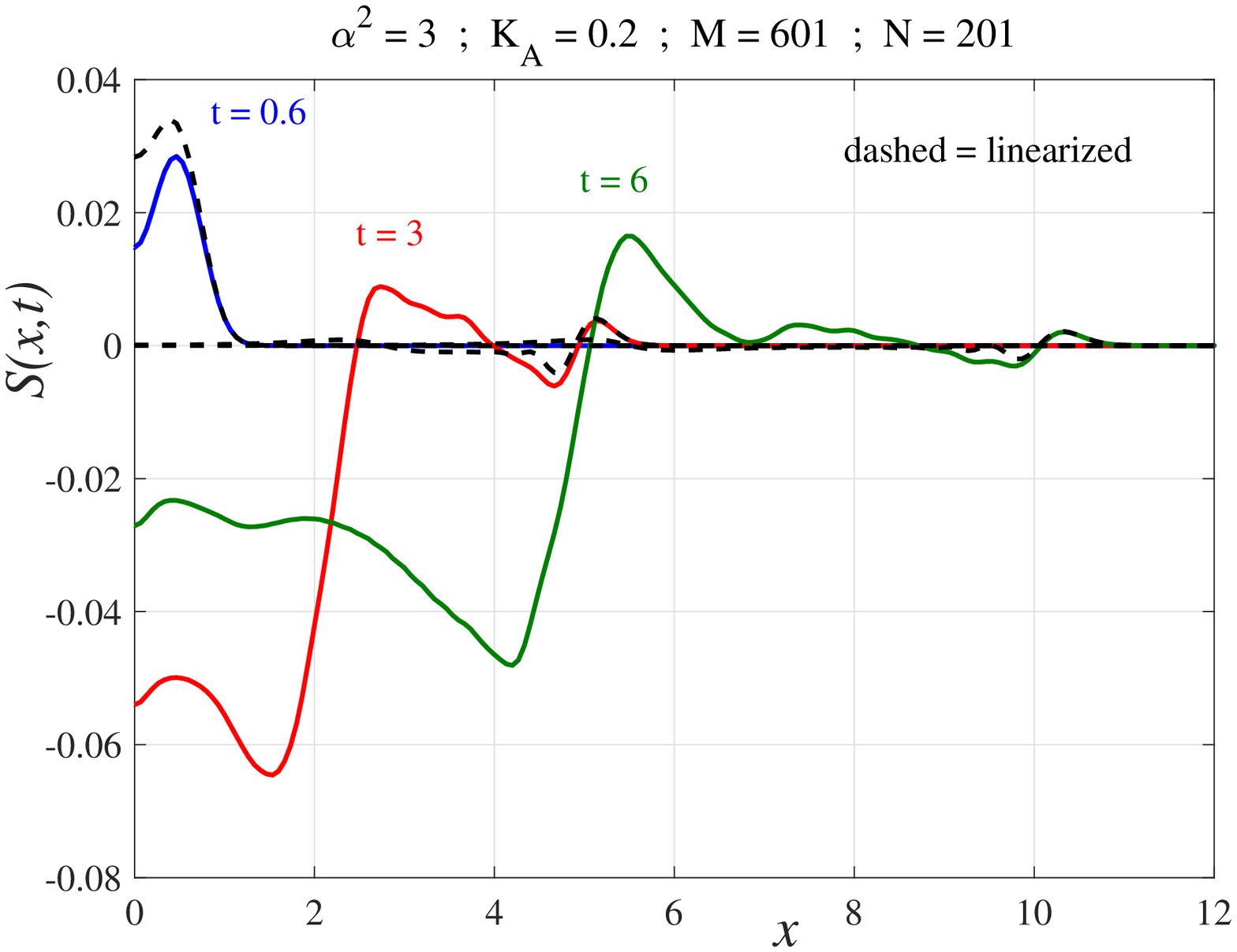}
\includegraphics[width=0.6\textwidth]{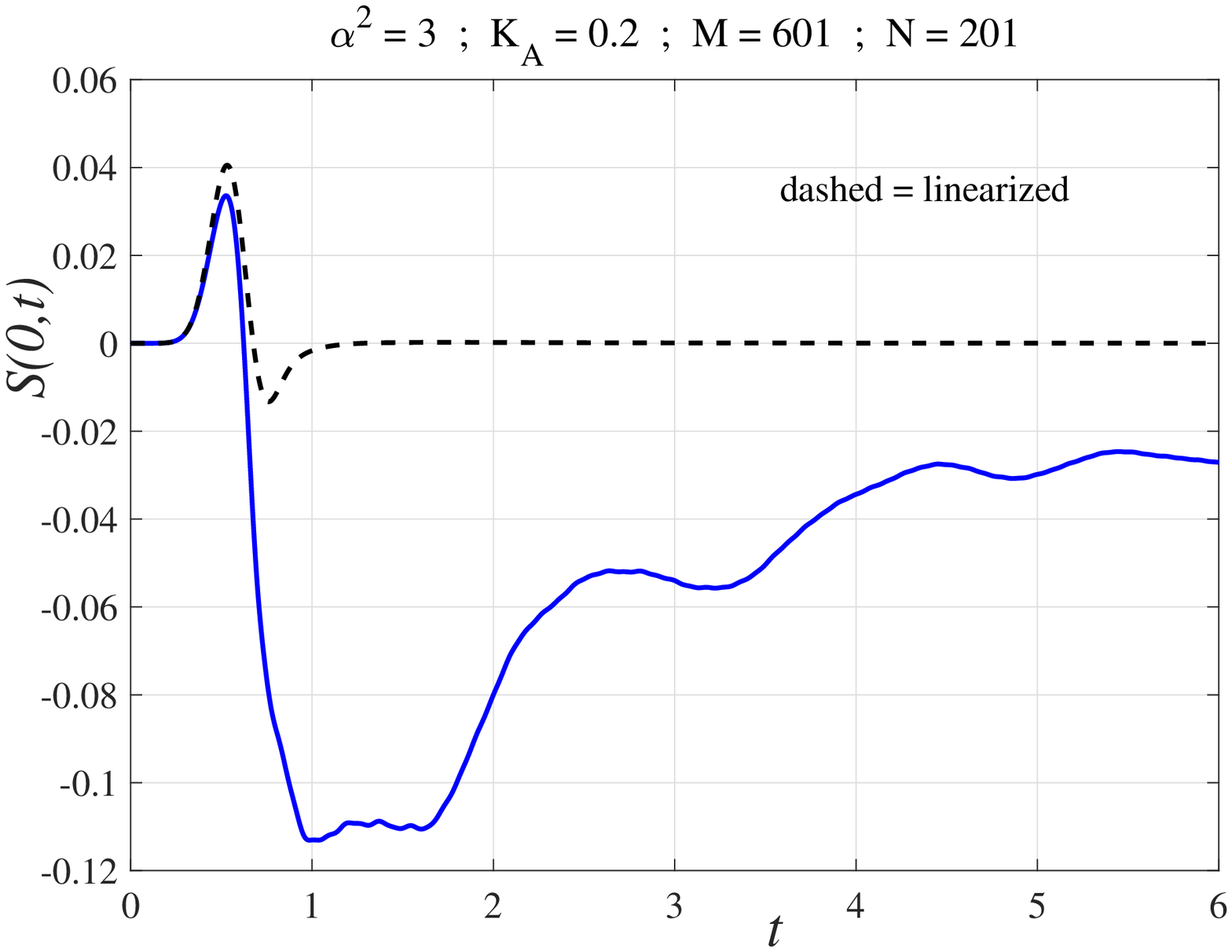}
\caption{A comparison of the linearized and nonlinear solutions for
speed ratio $\alpha^2 = 3$ and initial disturbance amplitude $K_A = 0.2$.  
(a) Surface profiles are shown for the three times $t = 0.6$, $3$ and $6$, and 
the linearized elevations are sketched with dashed lines;
(b) The time history of the surface elevation $S$ at the centre point $x = 0$.}
\label{fig_Figure4}
\end{figure}

A further set of wave profiles is shown in Figure \ref{fig_Figure4}(a) at this 
speed ratio $\alpha^2 = 3$.  Here, however, the initial disturbance amplitude has been
substantially increased again, now to the value $K_A = 0.2$.  The linearized 
solution, sketched with dashed lines, continues to over-estimate the height of the
bump that is formed at earlier times above the disturbance, as is again evident at
the earliest time $t = 0.6$ shown.  At this large amplitude, the nonlinear solution 
predicts a very substantial rebound near the location $x = 0$ of the initial
disturbance, although this feature is almost entirely absent from the linearized
approximation, which is predicated on the assumption that the interface remains
almost flat.  Nonlinear effects are therefore highly significant in this region,
and this is illustrated further in part (b).  In addition, the nonlinear free surface
exhibits several later waves of moderate amplitude, and a much slower return to the
neutral height $y = 0$.

\begin{figure}
\centering\includegraphics[width=0.8\textwidth]{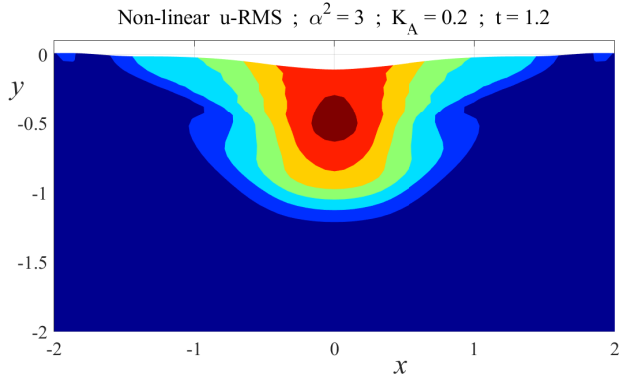}
\caption{A portion of the elastic domain near the free surface, for the
nonlinear solution obtained with $\alpha^2 = 3$ and $K_A = 0.2$ 
in Figure \ref{fig_Figure4}, at time $t = 1.2$.}
\label{fig_Figure5}
\end{figure}

From Figure \ref{fig_Figure4}(b), it is evident that, for the parameter values used
in that diagram, the maximum nonlinear disturbance to the free surface occurs at 
about time $t = 1.2$.  Accordingly, we show in Figure \ref{fig_Figure5} a portion 
of the elastic material near the origin.  Here, the different shadings (colours online) 
represent contours of the root mean squared elastic displacement
\begin{equation}
\| {\bf u} \| = u_{RMS} = \sqrt{ \left( u^X \right)^2 + \left( u^Y \right)^2 }
\label{eq:RMSdisplacement}
\end{equation}
in the material, deformed according to the nonlinear transformation (\ref{eq:NonDefineEta}).  
The depression at the free surface can be seen clearly, and there are also small secondary 
waves either side of it.

\begin{figure}
\centering
\includegraphics[width=0.6\textwidth]{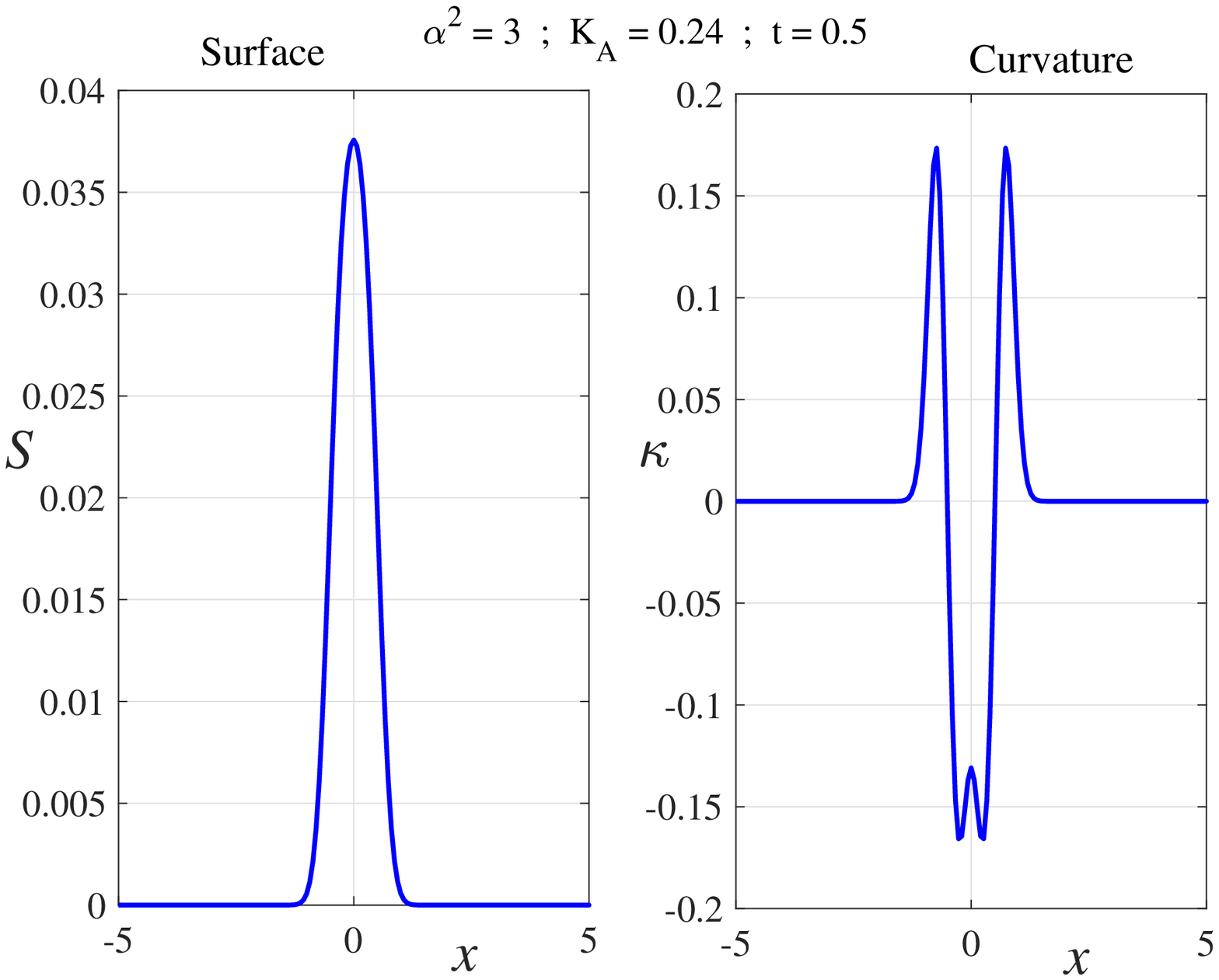}
\includegraphics[width=0.6\textwidth]{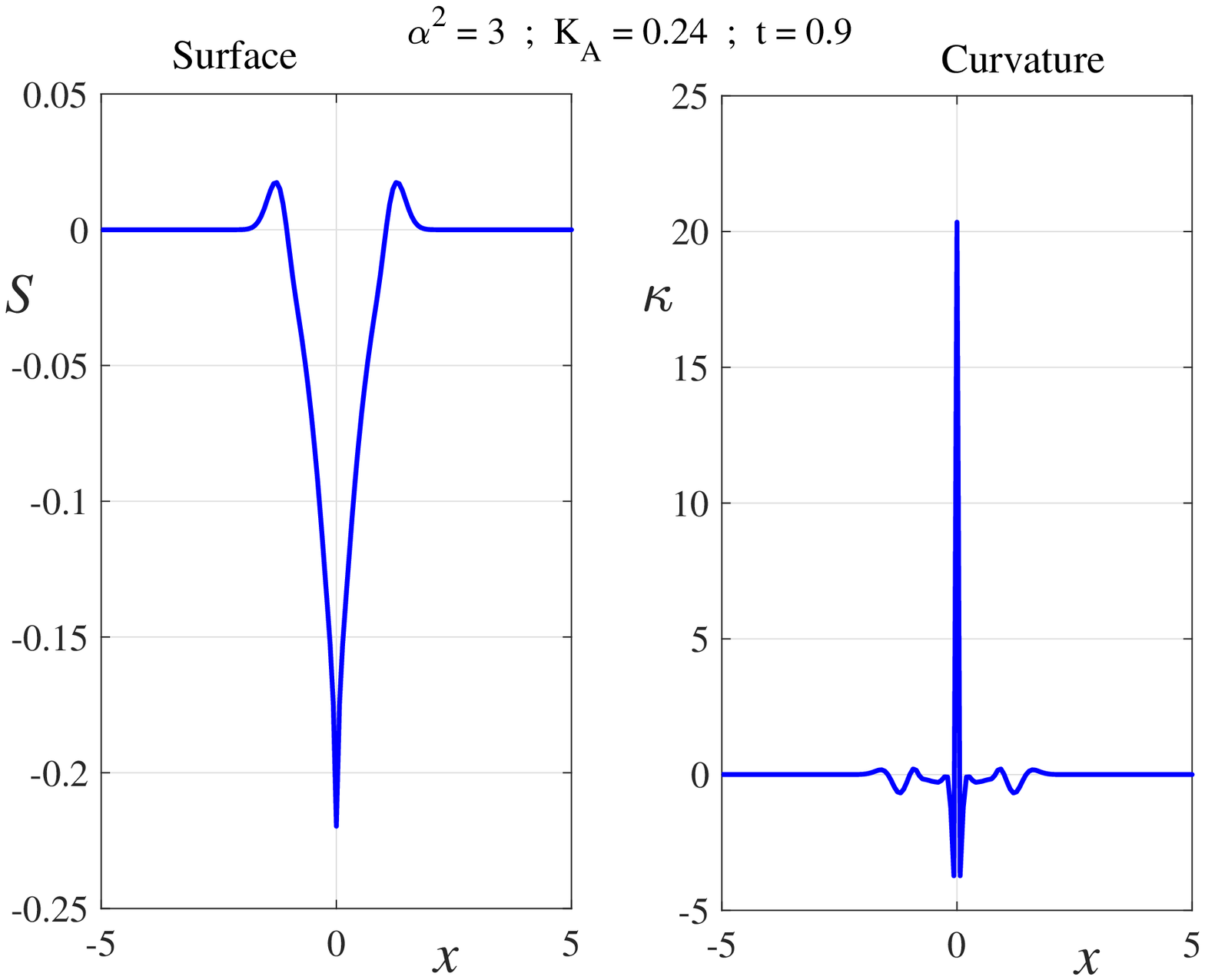}
\caption{The free-surface profile and its curvature at two different
times  (a) $t = 0.5$ and (b) $t = 0.9$, for speed ratio $\alpha^2 = 3$ 
and initial disturbance amplitude $K_A = 0.24$.}
\label{fig_Figure6}
\end{figure}

As the amplitude $K_A$ of the initial disturbance is increased, it is found that
the solution algorithm in Section \ref{sec:nonlinear} for the solution of the nonlinear
problem eventually fails at some finite time.  This is illustrated here for speed
ratio $\alpha^2 = 3$ in Figure \ref{fig_Figure6}, for the large value $K_A = 0.24$
of the submerged disturbance.  Solutions are shown at the two times $t = 0.5$
and $0.9$, in parts (a) and (b) respectively (computed with smaller time step
$\Delta t = 0.0001$), and the second of these two times is very close to the time 
at which the solution algorithm fails.

In order to account for this solution failure at finite time, we have also
calculated the curvature $\kappa$ of the interface $y = S \left( x,t \right)$,
according to the usual formula
\begin{equation}
\kappa = \frac {S_{xx}}{ \left[ 1 + \left( S_x \right)^2 \right]}   .
\label{eq:curvature}
\end{equation}
In this formula, the subscripts again denote partial derivatives, and these
are calculated using second-order accurate centred differences in our
finite-difference scheme.

The curvature (\ref{eq:curvature}) has also been plotted in Figures \ref{fig_Figure6} 
(a) and (b), on the right in each diagram, and helps explain the failure of the solution 
after some time.  At first the surface is initially flat, $S \left( x,0 \right) = 0$,
and a hump then forms, centred at $x = 0$.  This is visible in part (a) at the time
$t = 0.5$.  Nonlinear effects are then again responsible for a strong rebound
near this region, as shown in part (b) at the later time $t = 0.9$.  At this
value $\alpha^2 = 3$ of the squared speed ratio, a large spike in curvature forms at
the trough of this rebound pulse, where the surface develops a sharp cusp-like 
structure.  These features are visible in Figure \ref{fig_Figure6} (b), where the profile
clearly resembles a ``peakon'' with discontinuous slopes at its trough, and we suggest that
it is likely to be associated with brittle failure of the material.  To explore material 
failure in more detail, however, would require the inclusion of some fracture criterion, 
which goes beyond the scope of the present investigation and has not yet been attempted.  
For sufficiently large initial disturbance amplitude $K_A$, we have encountered similar 
peakon profiles with their associated curvature spikes at finite time, for all the values 
of squared speed ratio $\alpha^2$ we have investigated, although not all occur directly 
above the disturbance, as happens in Figure \ref{fig_Figure6} (b).

\begin{figure}
\centering
\includegraphics[width=0.6\textwidth]{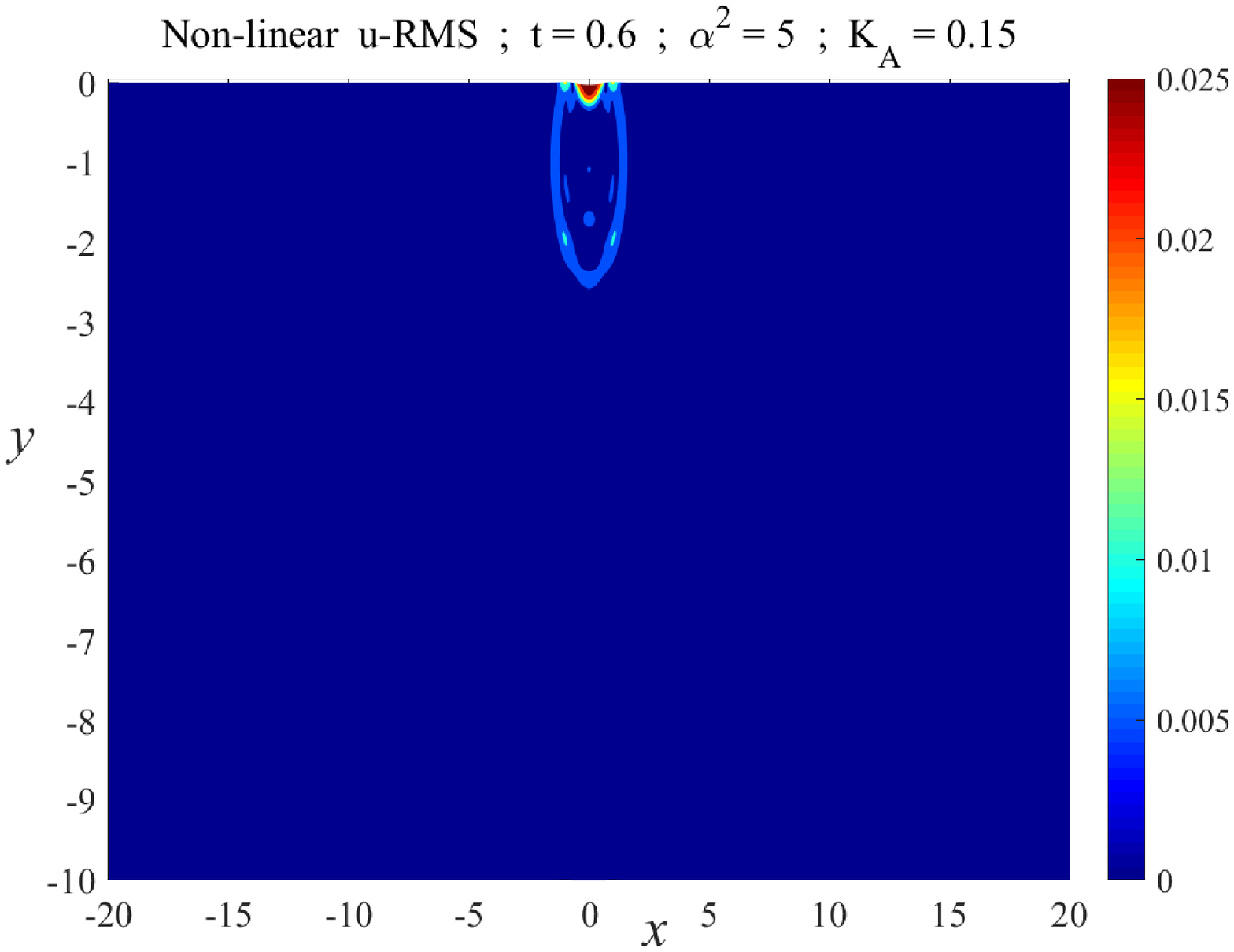}
\includegraphics[width=0.6\textwidth]{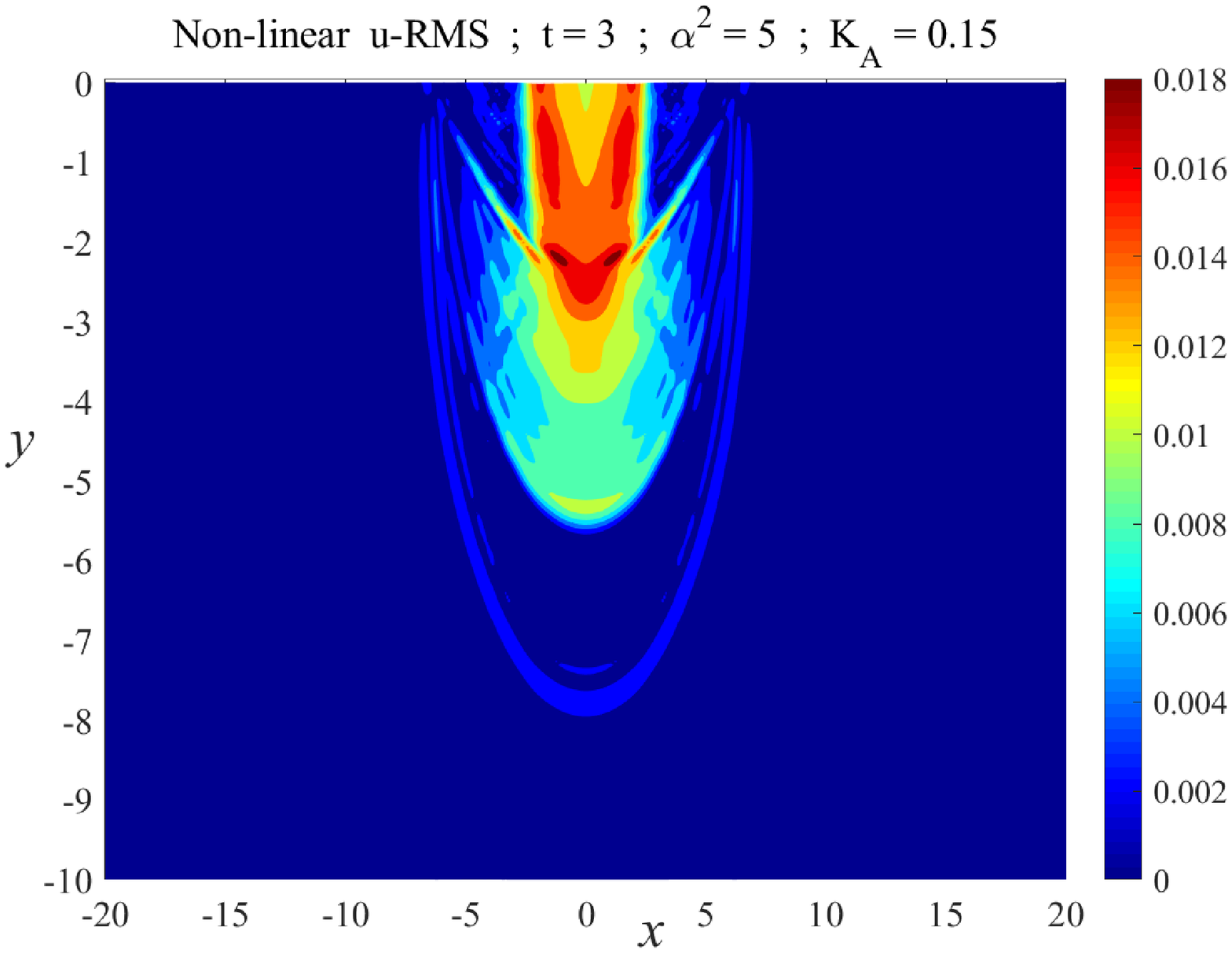}
\includegraphics[width=0.6\textwidth]{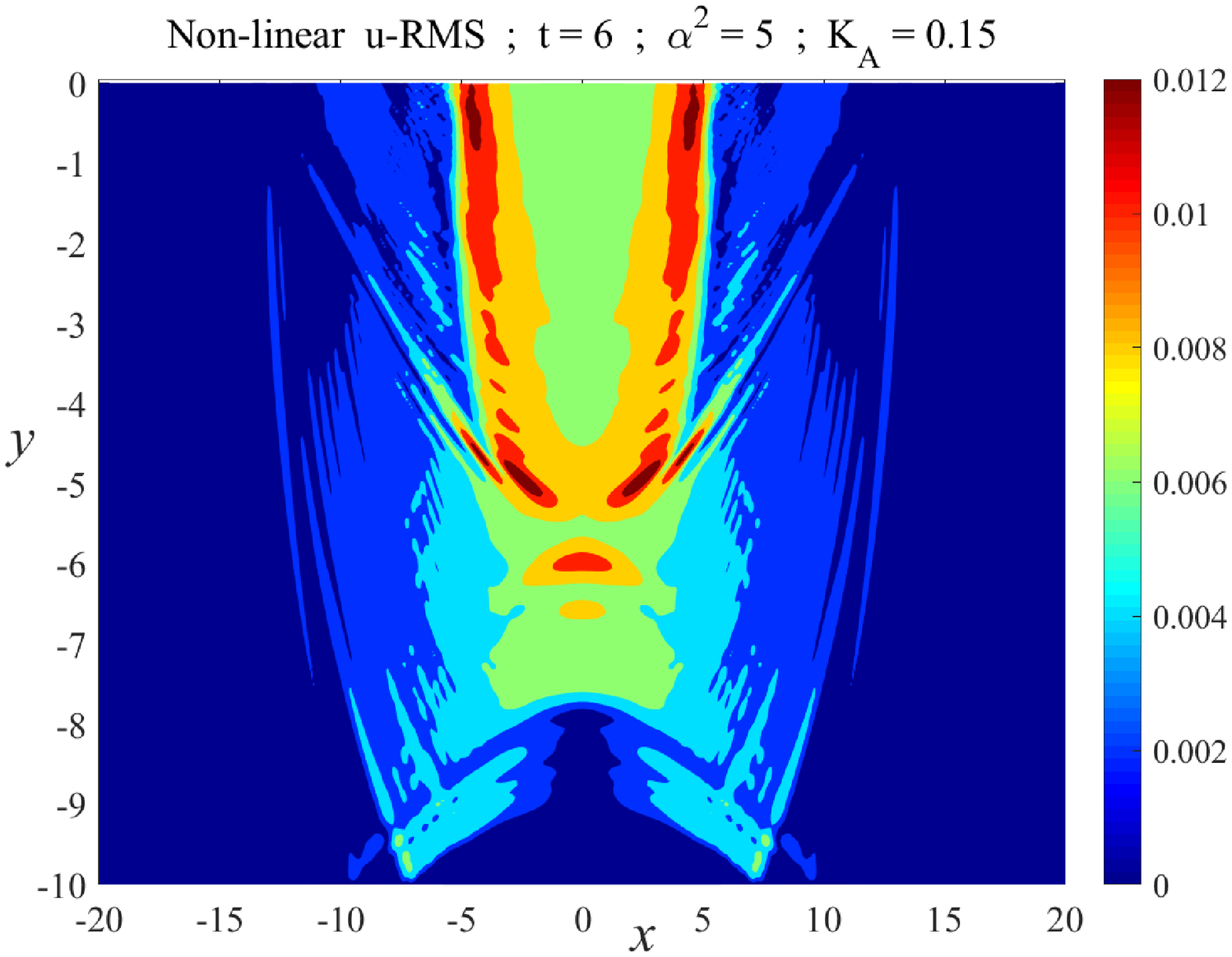}
\caption{The propagation of a disturbance within the elastic medium
at three different times  (a) $t = 0.6$, (b) $t = 3$ and (c) $t = 6$, 
for squared speed ratio $\alpha^2 = 5$ and initial disturbance amplitude $K_A = 0.15$.}
\label{fig_Figure7}
\end{figure}

The decay of the free-surface waves as time progresses is associated with the fact that
energy is also transmitted through the elastic medium.  This is illustrated here in
Figure \ref{fig_Figure7} at the three different times $t = 0.6$, $3$ and $6$, 
for the squared speed ratio $\alpha^2 = 5$ and initial disturbance amplitude
$K_A = 0.15$.  In these three diagrams, we have plotted contours of the root-mean squared 
displacement (\ref{eq:RMSdisplacement}) over the computational domain $-20 < x < 20$, 
$-10 < y < S \left( x,t \right)$ used here.  There are $M = 601$ and $N = 201$ points 
over this domain and the time step was reduced to $\Delta t = 0.001$.

At time $t= 0.6$ in part(a), the disturbance originally centred at
the point $\left( x,y \right) = \left( 0, -1 \right)$ has spread downwards and also 
up towards the surface, where it has begun to interact with the free surface and cause it
to deform.  It continues to spread both outwards and down in part (b), and this develops
in part (c).  In fact, the pulse has clearly reached the bottom of the computational
domain in part (c), at time $t = 6$, and it can be seen to have reflected off this
bottom in this last frame.  This will have no effect on the interface at this time,
however.

\section{Conclusions}
\label{sec:conclude}

Finite-difference methods have been developed in this work, for describing the
propagation of nonlinear seismic waves due to a submerged disturbance in an
elastic half space.  A novel predictor--corrector implicit ADI scheme has been 
created for this purpose.  In addition, a similar scheme has been developed for
the linearized counterpart of this problem.  The solutions we have presented
here have resulted from an initial disturbance that has left-right symmetry,
and they retain that symmetry throughout their evolution.  It is possible to
incorporate this bilateral symmetry into our numerical scheme, and indeed, 
this has been done for a version of the simple explicit method for the 
linearized problem, presented in Section \ref{sec:explicit}.  However,
it is onerous to incorporate in the non-linear ADI scheme of Section
\ref{sec:nonlinear}, and so has not been pursued here.

We have shown two pronounced effects of retaining nonlinear terms at the
free surface, and thus removing the assumption that the displacement of the
surface is negligibly small.  The first such effect is that finite-amplitude 
surface waves also involve a strong section of rebound at the free surface,
in the region above the initial disturbance.  This is almost completely
overlooked when the classical linearized free-surface conditions 
(\ref{eq:LinKinem})--(\ref{eq:LinDynTangent}) are assumed, but this 
rebound effect can be significantly greater than the initial upward pulse formed 
at the surface.

The second intriguing effect of surface nonlinearity is that, for sufficiently 
large initial disturbances, finite-amplitude free-surface waves can evolve into
peakons.  These have been suggested in weakly nonlinear theories \cite{XieEtAl},
and have now been confirmed here in the full geometrically nonlinear model,
apparently for the first time.  The peak in their free-surface profiles is
associated with a singularity in the curvature, and we suggest that the physical
consequences of such a singularity may include brittle fracture of the material.
Interestingly, curvature singularity formation within finite time is also known to
occur in large-amplitude free-surface waves in fluids, and was first established 
by Moore \cite{Moore}.  Later, Krasny \cite{Krasny} and also Forbes \cite{Forbes2009}, 
among many others, have demonstrated numerically that for water waves, this free-surface 
curvature singularity is associated with overturning of the surface.  

In the present paper, we have not considered the effects of material nonlinearity,
and have instead studied a model in which a generalized Hooke's law holds within the solid, 
so that nonlinear effects are purely geometric in origin.  However, if the disturbance is
of sufficiently large amplitude, then material nonlinearity may also need to be accounted
for, and a review on this topic is presented by Mihai and Goriely \cite{Mihai}, for
example.  The effects of both geometric and material nonlinearities on large-amplitude
free-surface waves, as well as the possible inclusion of fracture criteria within the
material itself, remain for future investigation.

\begin{acknowledgements}
This research was supported in part by Australian Research Council 
grant DP190100418.
\end{acknowledgements}

%
%

\end{document}